\documentclass[mnsc,nonblindrev]{informs3} 

\OneAndAHalfSpacedXI

\usepackage{natbib}
 \bibpunct[, ]{(}{)}{,}{a}{}{,}%
 \def\bibfont{\small}%
\TheoremsNumberedThrough     
\ECRepeatTheorems

\EquationsNumberedThrough    

\usepackage[ruled]{algorithm2e} 

\usepackage[utf8]{inputenc}

\usepackage{hhline}
\usepackage{multirow}

\usepackage{subcaption}

\usepackage{comment}

\usepackage{hyperref}
\hypersetup{colorlinks=true, citecolor=blue, linkcolor=blue, urlcolor=blue}

\usepackage{wrapfig}

\newcommand{\sm}{\textrm{-}}
\newcommand{\smi}{{\sm i}}

\newcommand{\vcg}{\text{VCG}}
\newcommand{\vcgn}{\text{VCGN}}

\renewcommand{\epsilon}{\varepsilon}

\newcommand{\eu}{\bar{u}}

\DeclareMathOperator*{\Exp}{\mathbb{E}}

\newboolean{releaseMode}
\newboolean{pageBreaksBetweenSections}
\newboolean{pageBreaksBetweenSubsections}

\setboolean{releaseMode}{true}
\setboolean{pageBreaksBetweenSections}{false}
\setboolean{pageBreaksBetweenSubsections}{false}

\ifthenelse{\boolean{releaseMode}}{
    \usepackage[colorinlistoftodos,prependcaption,disable]{todonotes}
    \newcommand{\para}[1]{}

}{ 
    \usepackage[colorinlistoftodos,prependcaption]{todonotes}
    \presetkeys{todonotes}{inline}{} 
    \newcommand{\para}[1]{\textbf{§ #1:}}

} 

\ifthenelse{\boolean{pageBreaksBetweenSubsections}}{
    \let\oldSubsection\subsection
    \renewcommand*{\subsection}{\newpage\oldSubsection}
    \setboolean{pageBreaksBetweenSections}{true}
}{} 

\ifthenelse{\boolean{pageBreaksBetweenSections}}{
    \let\oldSection\section
    \renewcommand*{\section}{\newpage\oldSection}
}{} 

\begin{document}



\RUNAUTHOR{Bosshard and Seuken}

\RUNTITLE{Cost of Simple Bidding}


\TITLE{The Cost of Simple Bidding\\in Combinatorial Auctions\footnotemark[2]}

\renewcommand*{\thefootnote}{\fnsymbol{footnote}}
\footnotetext[2]{Some of the ideas presented in this paper were also
described in a one-page abstract that was published in the conference
proceedings of the 22nd ACM Conference on Economics and Computation (EC'21) \citep{Bosshard2021CostEC}.}
\renewcommand{\thefootnote}{\arabic{footnote}}

\ARTICLEAUTHORS{%
\AUTHOR{Vitor Bosshard}
\AFF{Department of Informatics, University of Zurich, \EMAIL{bosshard@ifi.uzh.ch}}
\AUTHOR{Sven Seuken}
\AFF{Department of Informatics, University of Zurich and ETH AI Center, \EMAIL{seuken@ifi.uzh.ch}}
}


\ABSTRACT{%
We study a class of manipulations in combinatorial auctions where bidders fundamentally misrepresent what goods they are interested in.
Prior work has largely assumed that bidders only submit bids on their bundles of interest, which we call \emph{simple} bidding: strategizing over the bid amounts, but \emph{not} the bundle identities.
However, we show that there exists an entire class of auction instances for which simple bids are never optimal in Bayes-Nash equilibrium, always being strictly dominated by \emph{complex} bids (where bidders bid on goods they are not interested in). 
We show this result for the two most widely used auction mechanisms: \textit{first price} and \textit{VCG-nearest}.
We also explore the structural properties of the winner determination problem that cause this phenomenon, and we use the insights gained to investigate how impactful complex bidding may be.
We find that, in the worst case, a bidder's optimal complex bid may require bidding on an exponential number of bundles, even if the bidder is interested only in a single good.
Thus, this phenomenon can greatly impact the auction's outcome and should not be ignored by bidders and auction designers alike.
}%




\maketitle






\section{Introduction}

\para{Introduction to CAs}
Combinatorial auctions (CAs) are used in settings where multiple heterogeneous goods are being auctioned off to multiple bidders \citep{CramtonEtAl2006CombAuctions}. 
In contrast to simpler designs like simultaneous ascending auctions,  a CA allows bidders to express arbitrarily complex preferences over \emph{bundles} of goods, including substitutes and complements preferences. CAs are frequently used in practice, for example for the sale of industrial procurement contracts \citep{Sandholm2013largescale} or for selling TV ad slots \citep{goetzendorff2014core}.

One of the key challenges in CA design is to find a mechanism that provides bidders with good incentives. At first sight, the well-known VCG auction \citep{vickrey1961counterspeculation,clarke1971multipart,groves1973incentives} may look like a suitable mechanism, given that it is efficient and strategyproof. Unfortunately, VCG can produce very low or even zero revenue in domains with complements \citep{ausubel2006lovely}. Therefore,  the CA designs used in practice are typically non-strategyproof mechanisms.

 The \emph{first-price package auction} has been used in many different domains for over 20 years (e.g., to auction off bus routes \citep{cantillon2007combination}, to procure milk supplies in Chile \citep{epstein2004combinatorial}, and to auction off wireless spectrum (e.g., in Norway in 2013 and France in 2011 \citep{kokott2017combinatorial}).
 More recently, second-price style designs for CAs like the \emph{combinatorial clock auction (CCA)} \citep{ausubel2006clock} have gained traction. The CCA uses the \textit{VCG-nearest payment rule}, which (while not strategyproof) is designed to charge bidders payments that are in the core but also as close to VCG as possible \citep{DayMilgrom2008CoreSelectPackageAuctions,DayCramton2012Quadratic,DayRaghavan2007FairPayments}.
The CCA has been used for over 10 years in many different countries to conduct spectrum auctions, generating over \$15 Billion in revenues. It has also been used to auction off  licenses for offshore wind farms \citep{ausubel2011auction}.

\para{Complexity in CAs}
Of course, CAs are known to be ``complex'' --- at least in some respects. For example, solving the winner determination problem (i.e., finding the efficient allocation) is NP-hard \citep{Rothkopf1998Nphard}. If bidders have fully general preferences, then finding the efficient allocation may also require exponential communication between the auctioneer and the bidders in the worst case \citep{nisan2006communication}.
Given that VCG-nearest is not strategyproof, it is not surprising that there may be opportunities for bidders for \emph{bid shading, free-riding} or even \emph{overbidding} to optimize their utility \citep{ausubel2020core,lubin2018designing,beck2013incentives}. Finally, iterative mechanisms like the CCA offer additional pathways for manipulation \citep{janssen2016spiteful}.

\para{Design goals of CAs}
But these findings should not distract us from the fact that CAs were originally designed to make bidding in the auction simpler for the bidders. For example, \citet{Cramton2013SpectrumAuctionDesign} writes: \emph{``In contrast [to the SAA], the combinatorial clock auction has more complex rules, but the rules have been carefully constructed to make participation especially easy.''} We are interested in analyzing to what degree CAs truly make participation ``easy for bidders.''
We study this question formally for the two payment rules most commonly used in practice: first-price and VCG-nearest.

One aspect that has been largely overlooked in the literature so far are bidders' incentives for bidding on goods they have zero value for.\footnote{In lab experiments, \citet{scheffel2011experimental} have observed bidders engaging in ``extended bundling,'' i.e., bidding on bundles that include goods of no value to them. However, they attribute all of these actions to mistakes by bidders, and not to utility-increasing strategies.} 
The research community has commonly assumed that it is sufficient for bidders to submit bids on their bundles of interest to maximize their utility. \citet{Cramton2013SpectrumAuctionDesign} writes: \emph{``For the most part, the bidder can focus simply on determining its true preferences for packages that it can realistically expect to win.''} 
 Similarly, \citet{Milgrom2007PackageAuctionsAndExchanges} writes: \emph{``[...] these sealed-bid package designs require that bidders name prices for each package of interest to them.''}
However, we show that the assumption that bidders only bid on their bundles of interest (which we call \emph{simple} bidding) is a very significant restriction of the game-theoretic analysis.
When this artificial restriction is lifted and bidders are allowed to make bids on arbitrary bundles of goods (which we call \emph{complex} bidding), bidders may be able to significantly increase their utility, and the different strategies may considerably change the auction's equilibrium outcome.

The need for complex bidding cannot arise in full information settings, where all bidders know each others' values:
\citet{DayMilgrom2008CoreSelectPackageAuctions} showed that for any auction using a core-selecting payment rule, there always exist Nash equilibria in \emph{semi-sincere bids}, which are a form of simple bids where the bidder simply underreports her true value for each bundle by the same amount.
It is straightforward to extend these results to show that even outside Nash equilibria, a bidder always has a best response that is a semi-sincere bid, against any bid profile of the other bidders.
Thus, we will focus exclusively on the incomplete information (Bayesian) case.

\para{Conceptual Contribution}
Our main conceptual contribution is to provide an explanation of why and how complex bidding manipulations work (Section~\ref{sec:complexbiddingintuition}).
To provide some intuition, consider a simple first-price single-item auction. When submitting a bid, bidders must make a trade-off between winning more often vs.\ obtaining a higher profit when winning.
In Section~\ref{sec:example}, we provide a small but striking example to demonstrate that this logic does not need to apply when the strategy space becomes multi-dimensional.
In that case, a bidder can place a complex bid which can be interpreted as containing multiple ``conditional'' bids placed on the same bundle of goods, such that only the most beneficial of these bids becomes part of the winning allocation.
With complex bidding, a bidder can extract the highest possible profit, while still always winning her desired bundle when it is profitable.
We provide a computational interpretation of complex bidding which suggests that the culprit of this strategic vulnerability is the fact that the auctioneer makes an ex-ante commitment to use an efficient allocation rule, i.e., to solve the winner determination problem optimally.

\para{Main Result 1: complex BNEs in LLG}
We next show that this is a general and robust phenomenon (Section~\ref{sec:complexfamily}).
For this, we define a class of sealed-bid auction instances based on the well-known LLG domain \citep{ausubel2006lovely}, using either first price or VCG-nearest as the payment rule.
We prove that, in any Bayes-Nash equilibrium (BNE) of an auction instance in this class, at least one bidder must submit a complex bid; thus, BNEs in simple strategies cannot exist.

\para{Main Result 2: Exponential Separation}
Given this understanding of complex bidding, we then explore how impactful it can be for one bidder to deviate from a simple bid to a complex one, holding the other bidders' bid distributions fixed (Section~\ref{sec:mainresult}). We prove that the strategy space of a CA has enough room allowing bidders to make conditional bids that are ex-post optimal in each of an exponential number of world states that may materialize.
It follows that the amount of utility gain from complex bidding manipulations can be very significant in the worst case.
A bidder who is interested in a single good (and thus has the simplest type of valuation possible) may be incentivized to bid on exponentially many bundles of goods, and increase her utility by an exponential amount in the process.
Thus, even bidders who only have a localized interest in a few goods cannot avoid having to reason about the entire bundle space of the CA, contradicting the original design goals of CAs as described by \citet{Milgrom2007PackageAuctionsAndExchanges} and \citet{Cramton2013SpectrumAuctionDesign}.

We construct this exponential separation both for the first price and VCG-nearest payment rules.
This is especially surprising for the latter, given that VCG-nearest is a minimum revenue core-selecting rule and thus minimizes the total potential gains of bidders deviating from their bids \citep{DayRaghavan2007FairPayments}.
Our results imply that this joint minimization can still leave individual bidders with very high incentives to deviate, by setting their payment very high (equal to first price) for no good reason.

\para{Generality / Applicability}
Importantly, these incentives for bidders do \emph{not} arise from some special property of the mechanism, e.g., being iterative, only eliciting part of the bidders' preferences, or being restricted in some other way. Rather, the bidders make rational use of the large strategy space provided to them by a direct revelation mechanism, i.e., a sealed-bid CA with no limits on the number of bundles reported.
%


\section{Related Work}
\label{sec:relatedwork}

\para{High-dimensional manipulations}
The incentives of bidders in sealed-bid CAs have been extensively studied.
Among papers that study manipulations where bidders make use of the strategy space of an auction in unintended ways, the work by \citet{cantillon2007combination} stands out. They study a multi-item first price reverse auction where strategic price discrimination occurs. In this auction, a bidder wants to sell two goods for which she has linear costs (the cost of producing both goods is the sum of individual costs), but makes a combinatorial bid that offers a bundle discount, thus misrepresenting the relationship between the goods as being complementary.
They fit their model to data from London bus route auctions, finding that bidders indeed might be engaging in this manipulation deliberately.

\citeauthor{cantillon2007combination} further observe that the optimization problem the bidder faces in some combinatorial auctions is equivalent to the multi-product monopolist pricing problem \citep{mcafee1989multiproduct}.
Thus, the incentives to exploit the complexity of the strategy space can arise on both sides of the market.
Also from the auctioneer's perspective, the optimal mechanism design problem in CAs has recently been studied in terms of revenue guarantees for simple mechanisms and bounds on the menu size for obtaining near-optimal revenue \citep{babaioff2020simple,babaioff2021menu}.

\para{Need for BNEs, Analytical and computational BNE results}
In real-world CAs (e.g., spectrum auctions), bidders are typically much better informed about their own values than their competitors' \citep{milgrom2004putting}.
This motivates a Bayesian analysis of auctions, where bidders know their own value but only have distributional information regarding the other bidders’ values.
This is the standard model in the analytical literature, which has mostly focused on characterizing BNEs in small sealed-bid auction domains \citep{baranov2010exposure,Goeree2013OnTheImpossibilityOfCoreSelectingAuctions,ausubel2020core}.
There has also been a surge in the development of algorithms to find BNEs (or rather, $\epsilon$-BNEs) numerically \citep{Rabinovich2013ComputingBNEs,Bosshard2018nondecreasing,Bosshard2020JAIR,bichler2021learning}.
Such algorithms have been used to search through entire design spaces of auction rules for the most favorable mechanisms \citep{lubin2018designing}.
What these approaches have in common is that they assume (either explicitly or implicitly) that bidders only bid on their bundles of interest, and thus the complex bidding manipulations we describe in this paper fall outside their scope by definition.

\para{Beck\&Ott}
The work most related to our is that of \citet{beck2013incentives}.
They introduce a phenomenon called \emph{overbidding}, in which a bidder bids above her true value for at least some bundles.
This stands in contrast to our definition of complex bidding, where a bidder cannot make a bid on a bundle for which her marginal value (given any sub-bundle) is zero.
The two types of bid are similar in spirit, but not identical: a bid can be an overbid, complex, both or neither.
\citeauthor{beck2013incentives} construct an auction instance with  an artificial ``favored-bidder'' payment rule, in which the only BNE contains overbids.
Upon inspection, these bids also turn out to be complex, according to our terminology.
Unfortunately, \citeauthor{beck2013incentives} do not show any results under incomplete information with a payment rule actually used in practice, and they do not explore the extent to which their examples generalize.


\para{Simplicity and Complexity in Mechanism Design}
Our work is also related in spirit to a recent thread of research that aims to understand what makes mechanisms ``simple to play''  \citep{li2017obviously,pycia2019theory,borgers2019strategically}.
That research focuses on strategyproof mechanisms, whether or not they are \emph{obviously} strategyproof (OSP), and what differences in bidder behavior arise between OSP and non-OSP mechanisms.
In contrast, we apprach the same high-level question from a different angle: we only consider non-strategyproof mechanisms, and attempt to distinguish between simple manipulations, which can reasonably be expected from the bidders (e.g., bid shading), and complex ones, which should not (e.g., increasing the dimensionality of their bid).

\section{Preliminaries}
\label{sec:fm}

\subsection{Formal Model}


A combinatorial auction (CA) is a mechanism used to sell a set $M = \{1, 2, \ldots, m \}$ of goods to a set $N = \{1, 2, \ldots, n \}$ of bidders.
Each bidder $i$'s preferences over bundles are captured via the bidder's \emph{valuation} $v_i$. Specifically, for each bundle of goods $K \subseteq M$, we let $v_i(K) \in \mathbb{R}_{\geq 0}$ denote bidder $i$'s  value for bundle $K$. We normalize these values such that $v_i(\emptyset)= 0$.
We assume that valuations satisfy \emph{free disposal}, i.e., for any pair of bundles $K, K'$ it holds that: $K \subseteq K' \Rightarrow v_i(K) \leq v_i(K')$.

We consider one-shot sealed-bid auction formats.
Each bidder submits a bid $b_i$ to the auction, which is a (possibly non-truthful) declaration of her whole valuation.
The bid profile $b = (b_1, \ldots, b_n)$ is the vector of all bids from all bidders, and the bid profile of every bidder except bidder $i$ is denoted $b_{\smi}$.
We require bids to be submitted using the XOR bidding language \citep{Nisan2006BiddingLanguagesForCombinatorialAuctions}. Thus, each bid is a set containing zero or more \emph{atomic bids} $(K, \beta)$, where $K$ is a bundle and $\beta \in \mathbb{R}_{\geq 0}$ the \emph{reported value} for bundle $K$. Because of the free disposal assumption, the reported value of any bundle for which no atomic bid was submitted is implicitly inferred to be the maximum reported value of any of its sub-bundles, or zero if there exist no such sub-bundles.
To simplify the notation, we write an XOR bid $\{(K, \beta),\ldots, (K', \beta')\}$ as $(K, \beta) \oplus \ldots \oplus (K', \beta')$.
Furthermore, with a slight abuse of notation, we let $b_i(K)$ denote the value $\beta$ which bidder $i$ has explicitly (or implicitly) reported for bundle $K$.
%


The CA has an allocation rule $X$, determining an allocation $x=X(b)$, where $x_i = X_i(b)$ denotes the bundle assigned to bidder $i$. 
We require the resulting allocation to be \emph{feasible},  i.e., $\forall i,j \in N: x_i \cap x_j = \emptyset$, and we let $\mathcal{A}$ denote the set of all feasible allocations. We only consider  \emph{efficient} allocation rules, i.e., rules that produce an allocation that maximizes \emph{reported social welfare} (which is the sum of bidders' reported values).\footnote{
When multiple allocations are efficient, we break ties in favor of having a larger number of winners, and in lexicographical order among allocations with the same number of winners.
However, a bidder is never allocated a bundle of goods for which her bid is 0, even if that means that the auctioneer keeps some of the goods.
}
The CA also has a payment rule $p$ which is a function assigning a payment $p_i(b) \in \mathbb{R}_{\geq 0}$ to each bidder $i$.
Together, the allocation $x$ and the payment vector $p$ are called the auction \emph{outcome}.

Given a bidder's valuation $v_i$ and a bid profile $b$ of all bidders, we let $u_i(v_i, b)$ denote bidder $i$'s utility obtained from the auction's outcome.
We assume that the utility function is \emph{quasi-linear} of the form $u_i(v_i, b) = v_i(X_i(b)) - p_i(b)$.
\subsection{Payment Rules and the Core}
In this paper, we analyze two  CA payment rules: \emph{first-price} and \emph{VCG-nearest}.

\begin{definition}
Given allocation rule $X$ and bid profile $b$, bidder $i$'s \emph{first-price} payment is:
\begin{equation}
    p_i(b) := b_i(X_i(b)).
\end{equation}
\end{definition}

To define VCG-nearest payments, we must first introduce VCG payments and the core. For this, we define the set of feasible allocations when only considering a set of goods $K \subseteq M$ as $\mathcal{A}_K$. We define the reported social welfare achieved by a subset $L \subseteq N$ of bidders when only using goods $K \subseteq M$ as follows:
 
\begin{equation}
    W(b,L,K) := \max_{x \in \mathcal{A}_K} \sum_{j \in L} b_j(x_j).
\end{equation}
Using this notation, we can now define standard VCG payments. 

\begin{definition}
Given allocation rule $X$ and bid profile $b$, bidder $i$'s \emph{VCG payment} is:
\begin{equation}
    \vcg_i(b) := W(b,N\setminus \{i\},M) - W(b,N\setminus \{i\}, M\setminus X_i(b)).
    \label{eq:vcg}
\end{equation}
In words, $i$'s payment is the reported social welfare achievable by all other bidders if $i$ was not present, minus the welfare they achieve in the presence of $i$.
\end{definition}

Note that we define bidder $i$'s VCG payments based on the bid profile $b$ and not based on the valuation profile $v$. 
This allows us to compute what the VCG payments \emph{would have been}, even if a payment rule other than VCG is used in the auction itself.
Furthermore, note that $\vcg_i(b)$ only depends on bidder $i$'s bid $b_i$ via $i$'s allocated bundle  $K = X_i(b)$. We can thus equivalently write the VCG payment as $\vcg_i(b_\smi, K)$.
This is convenient when we later interpret this quantity as $i$'s \textit{winning threshold} for bundle $K$, i.e., for bidder $i$, $\vcg_i(b_\smi, K)$ is the minimum bid she must submit on that bundle to win it, regardless of the payment rule.
\smallskip

\paragraph{The Core.}
Informally, a payment vector $p$ is said to be \emph{outside the} \emph{core} if a coalition of bidders is willing to pay more for the goods than what the mechanism currently charges the winners. To avoid such outcomes, Day and Milgrom (2008) introduced the idea of \emph{core-selecting payment} rules that restrict payments to be in the \emph{revealed core} (i.e., the core with respect to reported values). 
\begin{definition}
Given allocation rule $X$ and a bid profile $b$, an auction outcome is in the \emph{revealed core} if $X$ is efficient, the payments $p(b)$ are individually rational, i.e., $\forall i \, \, p_i(b) \leq b_i(X_i(b))$, and the following additional core constraints are satisfied:
\begin{equation}
    \forall L \subseteq N:
    \sum_{j \in N \setminus L} p_j(b) \geq W(b,L,M) - W(b,L, X_{L}(b)),
    \label{eq:core}
\end{equation}
where $X_L$ is the set of goods collectively allocated to the set of bidders $L$.\footnote{To provide some intuition for these core constraints, observe that each set of bidders $L$ can potentially form a \emph{blocking coalition}, and it imposes the constraint that the joint payment of the remaining bidders $N \setminus L$ must be at least as large as the joint loss of welfare incurred by $L$ due to the presence of $N \setminus L$.}
A payment rule is called \emph{core-selecting} if it always produces outcomes in the revealed core.
\end{definition}
Note that the first-price payment rule is core-selecting, because Equation~\eqref{eq:core} always holds.\footnote{
This is because 
$\sum_{j \in N \setminus L} p_j(b) + W(b,L, X_{L}(b)) = W(b,N, M) \geq W(b,L, M)$.
}
%
Among all payment vectors in the core, the \emph{minimum revenue core (MRC)} is the set of payment vectors that minimize the sum of the  payments of all bidders.
We are now ready to define the second payment rule we study in this paper.

\begin{definition}
Given an allocation rule $X$ and a bid profile $b$, the \emph{VCG-nearest} \textit{payment rule} computes the vector of payments in the minimum revenue core that minimizes the $L_2$ distance to the VCG payment vector (computed based on bid profile $b$).
\end{definition}
\subsection{Simplicity of Bids and Valuations}
\label{sec:fm_simplicity}

In this work, we are primarily interested in a bidder's incentive to deviate from a ``simple'' subset of the strategy space and make use of more elaborate strategies outside this subset.

We call a bid \emph{simple} if it only contains atomic bids on those bundles for which the bidder has a genuine interest in every good, which means that the true value of the bundle cannot be explained by a strict subset of the goods within that bundle.
Formally, a simple bid can only contain an atomic bid on bundle $K$ if
\begin{equation}
    \forall K' \subset K : v_i(K') < v_i(K).
    \label{eq:fm_simple_criterion}
\end{equation}
%
Given a valuation $v_i$, we let $\mathcal{B}^s(v_i)$ denote the \emph{set of simple XOR bids}, and we let $\mathcal{B}^c$ denote the set of all possible (complex) XOR bids.
Note that $\mathcal{B}^c$ does not depend on $v_i$, and always contains all of $\mathcal{B}^s(v_i)$.
When clear from context, we sometimes use the term complex bid to refer to non-simple bids exclusively, i.e., $\mathcal{B}^c \setminus \mathcal{B}^s(v_i)$.\footnote{
For the purpose of a precise definition, it is desirable that the set of complex bids includes all simple bids as well, because optimal strategies will often use both simple and complex bids simultaneously.}

\para{Overbidding vs complex bidding}
\begin{remark}
\label{rem:BroadnessOfOurDefinition}
Some readers may wonder why our definition of complex bidding is not broader, e.g., by requiring that simple bids not only disallow bids on bundles that violate \eqref{eq:fm_simple_criterion}, but also bids implying a valuation for a bundle $K$ that is any amount above the true marginal valuation of that bundle, i.e., $\exists K' \subseteq K$ s.t. $b_i(K) - b_i(K') > v_i(K) - v_i(K')$.
Many of our results could indeed be adapted to hold under such an expanded definition.
However, it is surprisingly difficult to broaden the definition of complex bidding without making it \emph{too} broad and capturing some bids that are strategically very straightforward as well. We provide an example and further discussion of this issue in the electronic companion (Section~\ref{app:overbiddingexample}).
Our definition ensures that it only includes bids that are highly unnatural, and thus makes it easier to crisply identify the phenomenon we study in this paper.
\end{remark}

\subsection{Expected Utility and Best Responses}

We consider CAs in a Bayesian setting where bidder $i$ knows her own valuation $v_i$, but she only has probabilistic information (i.e., a prior) over all other bidders' bids, represented by the random variable $B_\smi$.
Given this uncertainty, the goal of bidder $i$ is to submit a bid that maximizes her \emph{expected utility} $\eu_i$, which is defined as
\begin{equation}
    \eu_i(v_{i},b_i,B_\smi) := \Exp_{B_{\smi}} \left[ u_i(v_i, b_i, B_\smi) \right].
    \label{eq:util}
\end{equation}

We call the highest possible expected utility that can be achieved with any bid in $\mathcal{B}^s(v_i)$ the \emph{simple best response utility}, given by
\begin{equation}
    \eu_i^s(v_i,B_\smi) := \sup_{b_i \in \mathcal{B}^s(v_i)} \eu_i(v_i, b_i,B_\smi).
\end{equation}
A bid is a \emph{simple best response} if it is in $\mathcal{B}^s(v_i)$ and achieves the simple best response utility.\footnote{We take the supremum over bids instead of the maximum, because the maximum might not exist due to discontinuities in the utility function. In that case, a best response is technically the limit of a series of bids.}
Analogously, we call the highest possible expected utility that can be achieved with any bid in $\mathcal{B}^c$ the \emph{complex best response utility} or just the \emph{best response utility}, given by
\begin{equation}
    \eu_i^c(v_i,B_\smi) := \sup_{b_i \in \mathcal{B}^c} \eu_i(v_i, b_i,B_\smi).
\end{equation}
A bid is a \emph{complex best response} if it achieves the complex best response utility.
\subsection{Bayes-Nash Equilibrium}

In Sections~\ref{sec:complexbiddingintuition} and~\ref{sec:complexfamily}, we will analyze the behavior of bidders in equilibrium.
For this, we consider the situation where each bidder's valuation is a random variable $V_i$ with a probability distribution that is common knowledge.
Bidder $i$ is the only one to know $V_i$'s realization $v_i$, and applies a strategy $s_i$ to derive a bid $b_i = s_i(v_i)$, where $s_i$ is a function mapping valuations to bids and is also common knowledge.
We can then define a (complex) Bayes-Nash equilibrium in the standard way:
\begin{definition}
A \emph{Bayes-Nash equilibrium (BNE)} is a strategy profile $s^* = (s^*_1, \ldots, s^*_n)$ for which it holds that
\begin{equation}
\forall i, \forall v_i  :  \eu_i(v_i,s^*_i(v_i),s^*_\smi(V_\smi)) = \eu_i^{c}(v_i,s^*_\smi(V_\smi)).
\end{equation}
\end{definition}

For bidders who only play simple bids, we can define an analogous equilibrium concept where strategies must map all valuations to simple bids:
\begin{definition}
A strategy $s_i$ is \emph{simple} if 
\begin{equation}
    \forall v_i: s_i(v_i) \in \mathcal{B}^s(v_i).
\end{equation}
\end{definition}
\begin{definition}
A \emph{simple BNE} is a strategy profile $s^*$ consisting only of simple strategies and for which it holds that
\begin{align}
\forall i, \forall v_i  &:  \eu_i(v_i,s^*_i(v_i),s^*_\smi(V_\smi)) = \eu_i^s(v_i,s^*_\smi(V_\smi)).
\end{align}
\end{definition}



\section{What Drives Complex Bidding under Incomplete Information?}

\label{sec:complexbiddingintuition}


\para{Overview and Informational Assumptions}
In this section, we provide a detailed structural analysis of why complex bidding is often incentivized in CAs.
As mentioned above, when bidders have full information about each others' valuations, bidder $i$'s set of best responses always includes at least one simple bid.
Thus, complex bidding can only be useful to a bidder under incomplete information.

We will show that complex bidding can help the bidder deal with uncertainty, increasing her profit while decreasing the risk of not winning her desired goods.
To strengthen this intuition, we first present an example where, in Bayes-Nash equilibrium, a complex bid allows one of the bidders to obtain a much better result than any simple bid.
We then take a single-agent perspective and look at the optimization problem that a given bidder faces, to help us understand the structural effects that create the incentives for complex bidding.

\subsection{Example}
\label{sec:example}

Consider the following auction, which is a version of a standard setting known as LLG (local-local-global), using the VCG-nearest payment rule.
To simplify the exposition, we use very simple, discrete value distributions for the bidders.
In Section~\ref{sec:complexfamily}, we show that our findings apply to much richer distributions as well.
\begin{example}
\label{ex:1}
We have two local bidders and one global bidder, bidding on two goods.
The global bidder has value $1$ for the bundle $\{1,2\}$ but $0$ for any single good.
It is known that the global bidder has a dominant strategy to bid truthfully \citep{beck2013incentives}.
The local bidders are interested in the singleton bundles $\{1\}$ and $\{2\}$, respectively, and have a 50\% chance of having either a low or high value, independently of each other.
The detailed valuations are given in Table~\ref{tab:example}, and the BNEs listed there are derived in the electronic companion (Section~\ref{app:BNEproof}).
\end{example}

\begin{table}
\TABLE
{ 
Auction instance where a complex strategy is better than all simple strategies for bidder 1. 
Bidders 1 and 2 have a 50\% chance (independently) to have either their low or their high value.
\label{tab:example}
}
{
\begin{tabular}{l|lr|ll|ll}
\bf Bidder & \multicolumn{2}{c|}{\bf Valuation} & \multicolumn{2}{c|}{\bf Simple BNE} & \multicolumn{2}{c}{\bf Complex BNE} \\
\bf ID &&& \bf Bid & \bf Exp.util. & \bf Bid & \bf Exp.util. \\

\hhline{=======}

1 & low \hspace{-1em}  & $v_1(\{1\}) = 0$ & $\emptyset$ & 0 & $\emptyset$ & 0\\
1 & high \hspace{-1em} & $v_1(\{1\}) = \frac{6}{5}$ & $(\{1\},\frac{1}{2})$ &0.35& $(\{1\},\frac{1}{2}) \oplus (\{1,2\}, 1)$ & 0.45 \\
\hline

2 & low \hspace{-1em} & $v_2(\{2\}) = 0$ & $\emptyset$ &0& $\emptyset$ &0 \\
2 & high \hspace{-1em} & $v_2(\{2\}) = \frac{4}{5}$ & $(\{2\},\frac{1}{2})$ &0.15& $(\{2\},\frac{1}{2})$& 0.15\\
\hline

3 &      & $v_3(\{1,2\}) = 1$ & $(\{1,2\}, 1)$ &0.5& $(\{1,2\}, 1)$ & 0.375\\
\end{tabular}
}
{}
\end{table}

\para{BNE description}
In the simple BNE, the local bidders win 25\% of the time, namely when both their valuations are high.
In that case, they outbid the global bidder, and they split the core payment of 1 evenly between them, because they both bid $1/2$.
In the complex BNE, the locals still win and split the payment in the high-high case. But additionally, when bidder 1 has a high value and bidder 2 has a low value, bidder 1 wins  over bidder 3 with her bid on the global bundle (as we break ties in favor of bidder 1). Thus, bidder 1 receives her desired good and pays 1, which increases her total expected utility from $0.35$ to $0.45$.

\para{BNE bidder 1 situation}
So, why does bidder 1 not simply increase her bid to 1 under simple bidding? 
This would have two effects: \textit{increasing} her utility in the high-low case by allowing her to win (just as in the complex BNE), but \emph{decreasing} her utility in the high-high case, because her higher bid causes her payment to increase.\footnote{This is because,  in this domain, the VCG-nearest payments are monotone in the bidder's bid; we characterize them in the electronic companion (Section~\ref{app:familyVCGN}).}
Adding these two effects together, bidder 1 is better off when bidding only $1/2$.
As a result, the simple BNE allocates the goods inefficiently 25\% of the time: in the high-low case, bidder 3 receives the goods, even though bidder 1 has strictly higher value for them.
In contrast, the complex BNE always produces the efficient allocation, and also yields higher expected revenue for the auctioneer.\footnote{
Observe that, in this example, the same strategy profile would still be a BNE even if bidder 1 had a marginal valuation for good $2$ given good $1$ in the range $0 < v_1(\{1,2\}) - v_1(\{1\}) < 0.5$.
However, as previously noted in Remark~\ref{rem:BroadnessOfOurDefinition}, we deliberately choose to keep this example more narrow and focused, to better show the effect we are studying.}


We next investigate in more detail why and how Example~\ref{ex:1} works, and we show that its underlying logic is robust and thus applicable to many auction domains.
In fact, this example is just one auction instance from a whole class of auctions covered by Theorem~\ref{thm:familyVCGN}.
%
\subsection{Winning Thresholds and Signals}
\label{sec:thresholds}

We now take a look at the optimization problem that a single bidder faces in an auction when keeping the strategies of the other bidders fixed, and we interpret complex bidding in that context.

\para{Winning Thresholds}
Given bids $b_\smi$, every bundle $K$ has an associated VCG payment $\vcg_i(b_\smi, K)$, which bidder $i$ cannot affect with her bid.
The efficient allocation rule dictates that a bid can only become part of a winning allocation if the welfare this bid adds to the allocation is at least as large as the welfare that it would displace. The latter amount is exactly the VCG payment.
Thus, $\vcg_i(b_\smi, K)$ is the minimum amount that bidder $i$ must bid to win bundle $K$ (assuming she makes no bid on any other bundle), and bidder $i$'s payment must be at least this winning threshold for any core-selecting payment rule.\footnote{This is because VCG payments are part of the core constraints and thus a lower bound for any such payment rule.}
%

\para{Threshold Distribution}
In an incomplete information setting, the bids $b_\smi$ are uncertain and represented by the random variable $B_\smi$.
Thus, the winning thresholds $\vcg_i(B_\smi, K)$ are also random variables.
Since the distribution of $B_\smi$ is common knowledge, the distribution of $\vcg_i(B_\smi, K)$ has a CDF $F_i^K$ that is known to bidder $i$.
%

\begin{figure}
\FIGURE
{
\begin{subfigure}[t]{0.46\textwidth}
\includegraphics[width=0.7\textwidth]{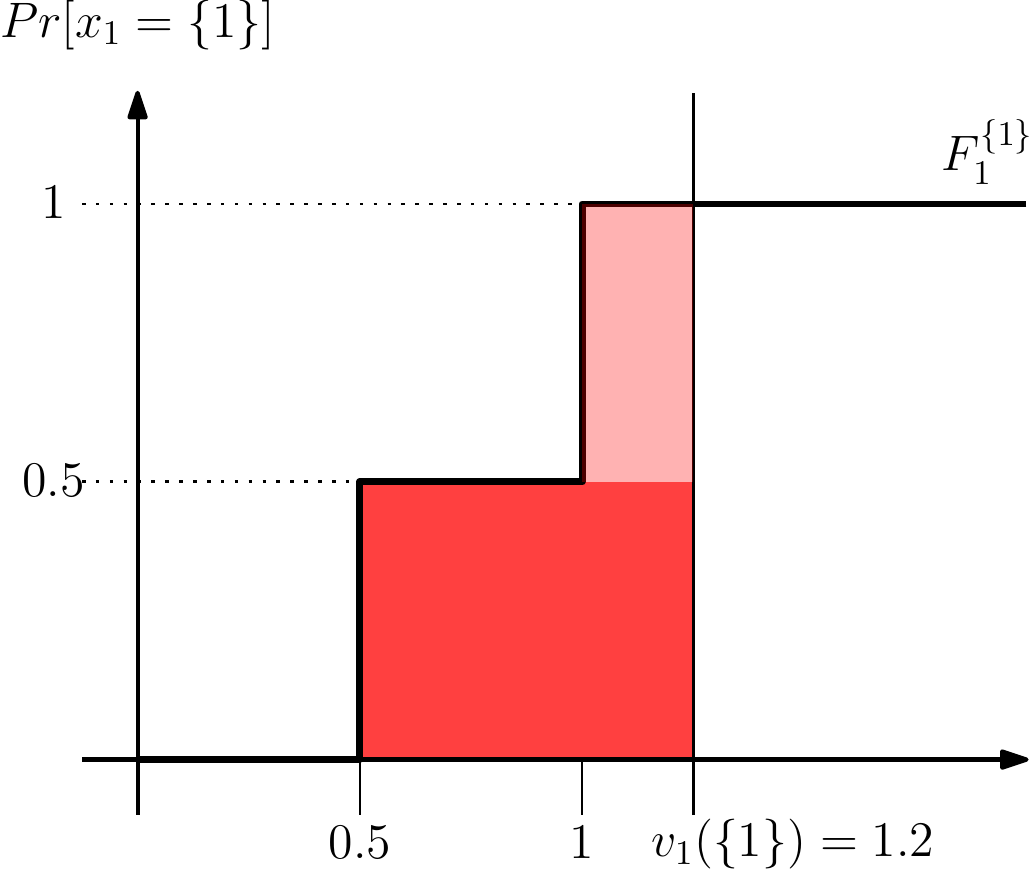}
\caption{$F_1^{\{1\}}$ from Example~\ref{ex:1}. The dark red slice corresponds to the event of probability $50$\% where $b_2(\{2\}) = 0.5$ and thus $\vcg_1(B_{\sm 1}, \{1\}) = 0.5$.}
\label{fig:area0}
\end{subfigure}
\qquad
%
\begin{subfigure}[t]{0.46\textwidth}
\includegraphics[width=0.7\textwidth]{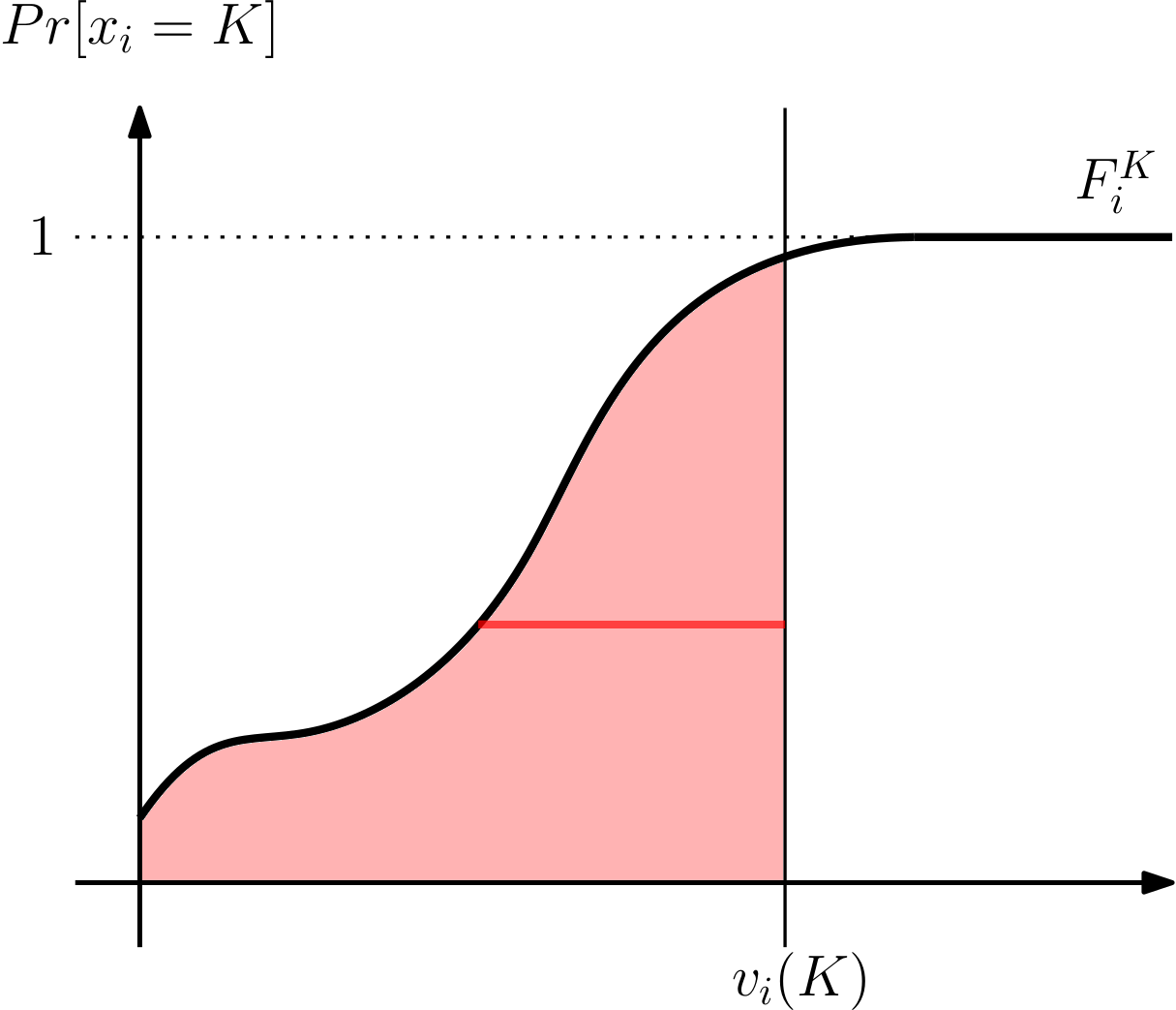}
\caption{Arbitrary $F_i^K$. The dark red (infinitesimal) slice is a single event with density $q > 0$.}
\label{fig:area2}
\end{subfigure}
}
{
Geometric intuition for bidder $i$'s utility under the VCG payment rule.
The total expected utility corresponds to the area under the curve of $F_i^K$ and to the left of $v_i(K)$.
\label{fig:area}
}
{}

\end{figure}

\para{Geometric interpretation}
Now observe that, for an auction with the VCG payment rule,  the utility obtained by a single-minded bidder who bids truthfully on bundle $K$ is simply the area under the graph of $F_i^K$.
To illustrate this geometrically, we return to Example~\ref{ex:1}, where the distribution $B_\smi$ is discrete and thus $F_i^K$ is a step function (Figure~\ref{fig:area0}).
%
%
%
%
The dark red area corresponds to the event of probability 50\% that bidder 2 has a high value, where the wining threshold for bidder 1 is 0.5.
The contribution of that event to bidder 1's expected utility under VCG is $0.5 \cdot (1.2 - 0.5) = 0.35$.
Likewise, the light red area corresponds to the event that bidder 2 has a low value, and the contribution to bidder 1's expected utility is $0.5 \cdot (1.2 - 1) = 0.1$. Bidder $1$'s total expected utility under VCG is the sum of these two terms, i.e., $0.35 + 0.1 = 0.45$.
This reasoning generalizes to continuous distributions, where we integrate over an infinite number of possible realizations of $\vcg_i(B_\smi, K)$ (Figure~\ref{fig:area2}).


\para{Complex Better Response}
For payment rules other than VCG, a given bid only captures part of the area under the curve.
For example, with the first price payment rule, the utility obtained in each realization of $B_\smi$ only spans from $b_i(K)$ to $v_i(K)$, and the utility between $\vcg_i(B_\smi, K)$ and $b_i(K)$ is lost.
When aggregating over all realizations of $B_\smi$, the expected utility obtained is a rectangle, corresponding to the multiplication of a specific \emph{payoff} $v_i(K) - b_i(K)$ targeted by the bid, times the \emph{probability} $F_i^K(b_i(K))$ of winning with that bid.

\begin{figure}
\FIGURE
{
\begin{subfigure}[t]{0.45\textwidth}
\includegraphics[width=\textwidth]{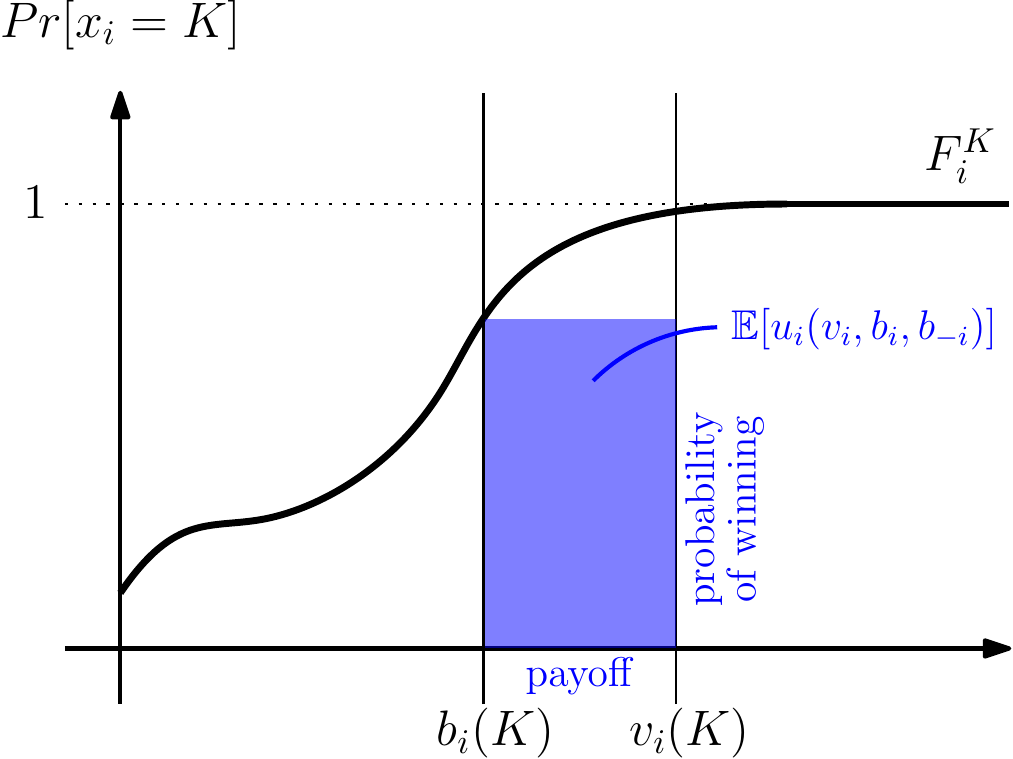}
\caption{Expected utility achievable with a simple bid.\\ \ }
\label{fig:geometric_interpretation_b}
\end{subfigure}
\hspace{3em}
\begin{subfigure}[t]{0.45\textwidth}
\includegraphics[width=\textwidth]{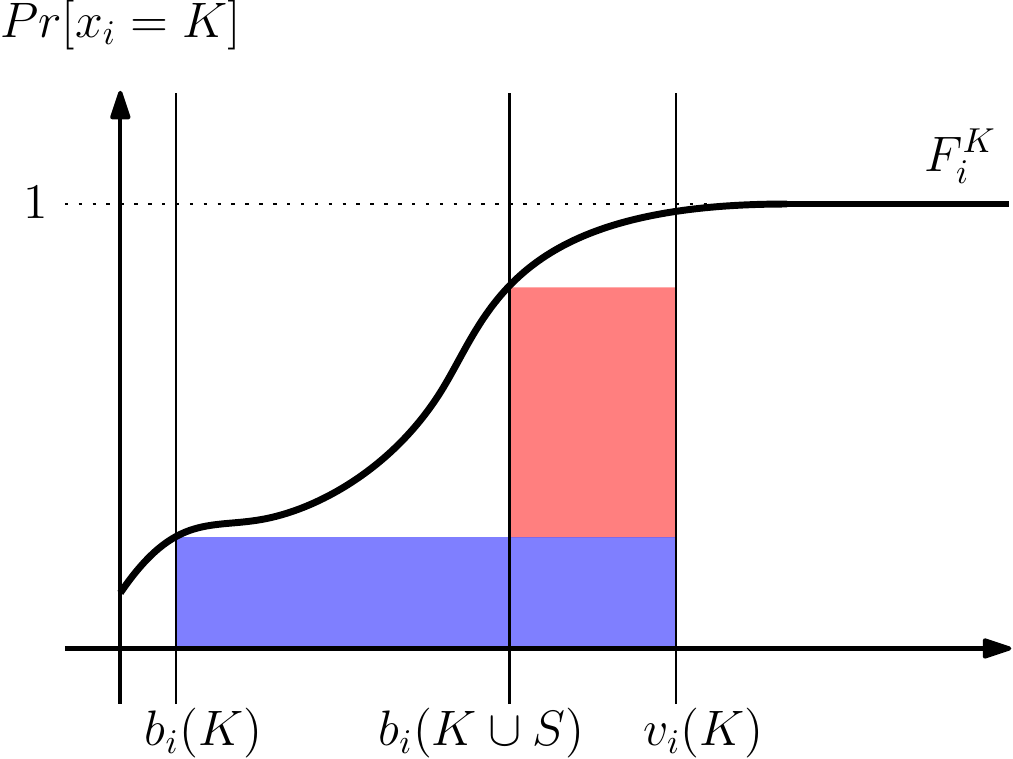}
\caption{Expected utility achievable with a correctly constructed complex bid.}
\label{fig:geometric_interpretation_c}
\end{subfigure}
}
{
Geometric intuition for bidder $i$'s utility optimization problem under the first price payment rule.
For simple bidding, the highest expected utility achievable corresponds to the largest rectangle fitting under the graph of $F_i^K$, to the left of $v_i(K)$. For complex bidding, the union of multiple such rectangles can be obtained.
}
{}

\end{figure}

Now, if bidder $i$ was only allowed to bid on bundle $K$, she would simply need to find the bid that yields the largest rectangle fitting under the graph of $F_i^K$ (Figure~\ref{fig:geometric_interpretation_b}), and her optimization problem would thus be reduced to a single dimension.
However, we claim that in the full strategy space of the auction, the bidder can place multiple atomic bids in such a way that the total utility captured is the union of the rectangles formed by multiple different bids (Figure~\ref{fig:geometric_interpretation_c}), which potentially captures much more utility than a single rectangle.


To see how this can be achieved, we first imagine a situation where bidder $i$ somehow managed to obtain information about the other bidders' bids $b_\smi$ from the auctioneer, before having to submit her own bid $b_i$.
It is clear that this information can be highly beneficial to bidder $i$, as it resolves all her uncertainty about $B_\smi$, allowing her to make a bid that is optimal ex-post: high enough to win, but not so high that she misses out on potential profit.
%
%

In our model, the bidder cannot actually obtain this information from the auctioneer directly, of course.
However, the auctioneer does guarantee that the efficient allocation rule will be used, and thus the distributions of the winning thresholds $\vcg_i(B_\smi, K)$ can be calculated in advance by bidder $i$, based on knowing the distribution of $B_\smi$.
Using this fact, bidder $i$ can place a carefully constructed complex bid that allows her to \emph{simulate} the actions she would take if she did have privileged information about $B_\smi$.
To see this, we need to consider the correlation structure between the random variables $\vcg_i(B\smi, K)$ and $B_\smi$.
Coming back to Example~\ref{ex:1}, it is easy to check that
\begin{align}
    \vcg_1(B_{\sm1}, \{1,2\}) &= 1,\\
    \vcg_1(B_{\sm1}, \{1\}) &= 1 - B_2(\{2\}),
\end{align}
which implies that
\begin{equation}
    \label{eq:correlationidentity} \vcg_1(B_{\sm1}, \{1,2\}) - \vcg_1(B_{\sm1}, \{1\}) = B_2(\{2\}).
\end{equation}
We highlight that $\vcg_1(B_{\sm1}, \{1,2\}) - \vcg_1(B_{\sm1}, \{1\})$ is \textit{inversely correlated} with $\vcg_1(B_{\sm1}, \{1\})$ itself, a feature that arises because the first two bidders must ``cooperate'' to win, i.e., the sum of their bids must jointly overcome the third bidder's bid for the goods.

Bidder 1 can exploit this correlation and place a bid $b_1$ that conditions on the realization of $B_2(\{2\})$ through the identity \eqref{eq:correlationidentity}, even though at the time of placing $b_1$, $B_2(\{2\})$ has not been realized yet from her perspective.
This is achieved by placing the XOR bid
\begin{equation}
    b_1 := (\{1\}, \beta) \oplus (\{1,2\}, \beta + t^*)
\end{equation}
for any value $t^* > 0$.
In the winner determination problem, bidder 1's second atomic bid will be more competitive than her first one exactly when
\begin{equation}
    b_1(\{1,2\}) - b_1(\{1\}) \geq \vcg_1(B_{\sm1}, \{1,2\}) - \vcg_1(B_{\sm1}, \{1\}),
\end{equation}
which due to \eqref{eq:correlationidentity} corresponds to
\begin{equation}
    (\beta + t^*) - \beta = t^* \geq B_2(\{2\}).
\end{equation}
Thus, bidder 1 always offers a low bid of $\beta$ for her desired good, but increases this bid by $t^*$ exactly when needed, namely when $b_2(\{2\}) = 0$ and she has to overcome bidder 3's bid by herself.
In the complex BNE, bidder 1 chooses $\beta$ and $t^*$ optimally, and as a result wins her desired bundle 100\% of the time, never paying more than VCG.

\para{generalization}
More generally, it is possible for bidder $i$ to condition on the realization of $B_\smi$ whenever she is interested in a bundle $K$, and there exists another bundle $S$ such that
\begin{equation}
    \vcg_i(B_\smi, K \cup S) - \vcg_i(B_\smi, K)
    \label{eq:marginaldemand}
\end{equation}
is inversely correlated with $\vcg_i(B_\smi, K)$.
We emphasize that such dynamics \textit{always} arise when some of the other bidders' bids make it easier (rather than harder) for bidder $i$ to become a winner.
This is a very common occurrence, and the LLG domain is just the smallest auction domain exhibiting this effect.


\begin{wrapfigure}{R}{0.5\textwidth}
\begin{minipage}{0.5\textwidth}
\vspace{-5mm}
\begin{algorithm}[H]
\SetAlgoLined
\DontPrintSemicolon

$signal := \vcg_i(B_\smi, K \cup S) - \vcg_i(B_\smi, K)$ \; 

\eIf{
    signal $< t^*$
}{
    $b_i = (K, \beta + t^*)$ \;
}{
    $b_i = (K, \beta)$ \;
}

\caption{Bidder $i$'s bid-program}
\label{alg:intuition}
\end{algorithm}
\vspace{-3mm}
\end{minipage}
\end{wrapfigure}

\para{Computational Interpretation}
A more computational way of thinking about this phenomenon is that bidder 1 does not bid a fixed numerical value, but (figuratively) provides a computer program as a bid (Algorithm~\ref{alg:intuition}).
This ``bid-program''  gets executed in an environment where the marginal demand for bundle $S$ given bundle $K$ as expressed in \eqref{eq:marginaldemand} is readily accessible, and the bid-program can use this information as a \emph{binary signal}, to select one of two possible atomic bids to be submitted to the auctioneer, and retracting the other.
In Section~\ref{sec:mainresult}, we will construct a more complicated signal that allows the bidder to adapt her bid to be optimal for each of an exponential number of winning thresholds, by placing exponentially many atomic bids.

The manipulation we have discussed is structurally unavoidable whenever some of the winning thresholds are negatively correlated in the way shown; the computation that exposes the signal \emph{must} be performed by any auctioneer who commits to always allocating the goods efficiently, because this commitment already fixes how the winning thresholds $\vcg_i(B_\smi, K)$ are calculated, regardless of the payment rule used in the auction.
%
 

\section{Complex Bidding Dominates Simple Bidding in BNE}
\label{sec:complexfamily}

\newcommand{\family}{LLG-A\xspace}
\newcommand{\familyVCGN}{LLG-B\xspace}

\para{Outline}
In the previous section, we have shown that complex bidding can arise in the presence of a signal, and we argued that such signals arise \textit{naturally} in CAs.
To substantiate this claim, we now introduce a class of auction domains for which simple bidding is always dominated by complex bidding, in the sense that any BNE which might exist cannot entirely consist of simple strategies.
We show this for the two most commonly used payment rules, first-price and VCG-nearest.

\para{Auction family}
The class of auction domains we define next is a generalization of Example~\ref{ex:1}.
It has the same combinatorial structure as the LLG domain:
there is a global bidder with a known valuation for a bundle of two goods, and two local bidders who must cooperate to jointly win the goods.
However, we impose a few special conditions on the bidders' valuations.
In particular, we require two features that incentivize complex bidding:
(1) a local bidder's valuation can be high enough to want to win even without the other local bidder, and 
(2) that other bidder's valuation can be low enough that the first bidder cannot always count on her help.

These conditions are fulfilled in many scenarios, e.g., when the two local bidders have value distributions that are i.i.d. uniform on the interval $[0,2]$, with the global bidder having a fixed value of $1$.
%
As we aim to impose as few restrictions as possible on bidders' value distributions, we formalize the conditions in a minimal way, such that they also apply to more involved domains where valuations are non-uniform, correlated with each other, and so on.

\para{Auction class FP (formal)}
\begin{definition}
An auction domain belongs to the \family class if
\begin{itemize}
\item There are $2$ goods and $3$ bidders.
\item For $i \in \{1,2\}$, bidder $i$ is ``local'', with a single-minded valuation on bundle $\{i\}$.
\item Bidder $3$ is ``global'', with a single-minded valuation on bundle $\{1, 2\}$, with $v_3(\{1,2\}) = 1$.
\item The following relations hold for the bidders' valuations:
    \begin{itemize}
    \item  Bidder 1 can have a high value (higher than bidder 3):
    \begin{equation}
        \label{complexfamily-val1} \Pr \left[v_1(\{1\}) > 1 \right] > 0
    \end{equation}
    \item When bidder 1 has high value, bidder 2 does not always have 0 value:
    \begin{equation}
        \label{complexfamily-v2_not_always_zero} \forall \beta > 1: \Pr[v_2(\{2\}) > 0 \, | \,  v_1(\{1\}) = \beta] > 0
    \end{equation}
    \item Bidder 3 can have the highest value for the goods:
    \begin{equation}
        \label{complexfamily-val4} \Pr \left[v_1(\{1\}) + v_2(\{2\}) < 1 \right] > 0
    \end{equation}
    \item When bidder 1 has high value, bidder 2 can have arbitrarily low value:
    \begin{equation}
        \label{complexfamily-val2} \forall \beta > 1, \forall \epsilon > 0 : \Pr \left[v_2(\{2\}) < \epsilon \, \middle| \,  v_1(\{1\}) = \beta \right] > 0
    \end{equation}
    
    \end{itemize}
\end{itemize}
\end{definition}

We now show our first main result, namely that for any auction instance in \family, in any BNE, there is at least one bidder who plays a complex strategy for some of her valuations.
Thus, BNEs consisting only of simple strategies do not exist.
\begin{theorem}
    In any auction domain of the \family class, when using the first price payment rule, every BNE $s^*$ contains at least one strategy $s_i^*$ that is not simple.
    \label{thm:familyFP}
\end{theorem}


\proof{Proof.}

\para{Proof Structure}
The proof is structured as follows.
We focus exclusively on the case where $v_1(\{1\}) > 1$, and we assume that we are at a strategy profile $s^*$ that is a simple BNE.
Then, given the conditions on the class \family, we show that the distributions $F_1^{\{1\}}$ and $F_1^{\{1,2\}}$ must have certain shapes.
Given these shapes, we construct a complex best response for bidder 1 that captures strictly more utility than $s^*_1(v_1)$, showing that $s^*$ is not a (complex) BNE.

\para{bidders don't always make empty bids}
First, note that the global bidder must make a deterministic bid, so we have that $b_{3} = \{(\{1,2\}, \bar{\beta})\}$ for some $\bar{\beta} \leq 1$.
Because we are in simple BNE, we must have $\bar{\beta} > 0$:
Under first price, bidding above one's value is always strictly dominated, so due to (\ref{complexfamily-val4}), the local bidders must sometimes jointly bid below 1.
Therefore, setting $\bar{\beta} = 1-\epsilon$ yields strictly positive utility to bidder 3 and thus dominates $\bar{\beta} = 0$, which always yields 0 utility due to our tiebreaking definition.

Similarly, because we are in simple BNE, bidder 2 cannot always bid zero:
in that case, bidder $1$'s simple best response would clearly be to bid $\bar{\beta}$ whenever her value is strictly above $\bar{\beta}$.
This in turn means that any bid by bidder 2 that is strictly larger than 0 wins with positive probability, and thus strictly dominates the zero bid whenever bidder 2's value is above zero, which happens with positive probability due to (\ref{complexfamily-v2_not_always_zero}).

Given these arguments, we claim that bidder 1's winning thresholds have the following shapes:
\begin{itemize}
\item $F_1^{\{1,2\}}$ is a step function going directly from 0 to 1 at $\bar{\beta}$.
\item $F_1^{\{1\}}$ also reaches 1 exactly at $\bar{\beta}$, but it is \emph{not} such a step function.
\end{itemize}
Regarding the shape of $F_1^{\{1,2\}}$, we argue as follows:
this function must be 0 below $\bar{\beta}$, because any bid from bidder $1$ in this region always loses to the global bidder's bid.
Above $\bar{\beta}$, if the probability of winning was lower than 1, it would imply that bidder $2$ bids strictly higher than $\bar{\beta}$ with positive probability, which cannot happen in a simple BNE under first price, as it is strictly dominated by reducing the bid to $\bar{\beta}$, which is already high enough to guarantee that bidder 2 always wins.

Regarding the shape of $F_1^{\{1\}}$, we argue as follows:
Approaching $\bar{\beta}$ from below, $F_1^{\{1\}}$ must at some point go strictly above 0, as we already established that bidder 2 has positive probability of making a strictly positive bid on bundle $\{2\}$, which makes it easier for $1$ to win.
However, $F_1^{\{1\}}$ must remain strictly below 1 until it reaches $\bar{\beta}$: 
if we had $F_1^{\{1\}}(\bar{\beta} - \epsilon) = 1$ for some $\epsilon > 0$, this would imply that bidder $2$ always bids at least $\epsilon$, which due to condition (\ref{complexfamily-val2}) is suboptimal under first price.\footnote{
To see this, we compare the utility obtained by a bid strictly above the true value with a truthful bid.
In realizations of $B_{\sm 2}$ where both bids would win, the latter bid obtains stricty more utility as it has a lower payment.
In cases where the latter bid is not high enough to win, the VCG payment (i.e., the winning threshold), is above the true value, and thus the utility obtained by the bidder under the former bid would be negative.
}

Finally, above $\bar{\beta}$, $F_1^{\{1\}}$ is clearly 1, because bidder $1$ is guaranteed to win even without any assistance from bidder $2$.

Having established the shape of the winning thresholds for bidder 1, we consider bidder 1's simple best response $b_1 = s_1(v_1) = (\{1\}, \beta)$, and we split the remainder of the proof into two cases depending on the value of $\beta$, and show that in both cases, there is a complex bid that is strictly better than $(\{1\}, \beta)$.

\textbf{Case 1} ($\beta < \bar{\beta}$):
In this case, the following complex bid is a better response: $b'_1 = (\{1\}, \beta + \epsilon) \oplus (\{1, 2\}, \bar{\beta} + \epsilon)$ for some $\epsilon$ with $0 < \epsilon < \bar{\beta} - \beta$.

We split the realizations of bidder 2's bid $B_2$ into four subcases, depending on what bidder $1$ wins.
First, with some probability $q_1$, bidder $1$ is allocated $\{1\}$ under $b_1$, and in that case she is also always allocated $\{1\}$ under $b'_1$.
This is because the two local bids jointly sum up to at least $\bar{\beta} + \epsilon$, and so are at least as competitive as bidder $1$'s second atomic bid, and we assume that ties are broken in favor of having more winners.
Second, third and fourth, if bidder $1$ would be allocated nothing under $b_1$, then under $b_1'$ she has three possible allocations, $\{1\}$, $\{1,2\}$ and $\emptyset$, with probabilities $q_2, q_3$ and $q_4$, respectively.
It follows that the expected utility of bidder $1$ under $b_1$ and $b'_1$ is
\begin{align}
    \eu_1(v_1, b_1, B_\smi) =&q_1 \cdot (v_1(\{1\}) - \beta)\\
    \eu_1(v_1, b_1', B_\smi) =&q_1 \cdot (v_1(\{1\}) - \beta - \epsilon) + q_2 \cdot (v_1(\{1\}) - \beta - \epsilon) + q_3 \cdot (v_1(\{1,2\}) - \bar{\beta} - \epsilon).
\end{align}
Subtracting the former from the latter, the difference in expected utility is
\begin{equation}
    - (q_1 + q_2 + q_3) \cdot \epsilon + \underbrace{q_2 \cdot (v_1(\{1\}) - \beta) + q_3 \cdot (v_1(\{1,2\}) - \bar{\beta})}_{> 0}.
    \label{eq:thm4.2_epsilon_limit}
\end{equation}
Observe that $q_1 < 1$ due to (\ref{complexfamily-val2}) and the fact that bidder 2 never bids strictly above her value.
We also have that $q_4 = 0$, because bidder 1's second atomic bid $(\{1, 2\}, \bar{\beta} + \epsilon)$ implies that any winning allocation must have bid at least $\bar{\beta} + \epsilon$. This is higher than bidder 3's bid by construction, and bidder 2's bid is also always at most $\bar{\beta}$, since this strictly dominates any higher bid in simple BNE under first price.
%
It follows that either $q_2 > 0$ or $q_3 > 0$, and thus \eqref{eq:thm4.2_epsilon_limit} is larger than $0$ for small enough $\epsilon$.

\textbf{Case 2} ($\beta \geq \bar{\beta}$):
We can assume WLOG that $\beta = \bar{\beta}$, because this is always strictly better for bidder $1$.
Since $F_1^{\{1\}}$ is not a step function, we can always find a number $0 < \beta' < \bar{\beta}$ such that the bid $b'_1 = (\{1\}, \beta')$ wins with positive probability.
Given this number $\beta'$, we claim that the XOR bid $b''_1 = (\{1\}, \beta') \oplus (\{1,2\}, \bar{\beta})$ obtains more utility than $b_1$.
%


Observe that bidder 1 always wins bundle $\{1\}$ under $b_1$.
Now, whenever bidder 1 wins nothing under $b_1'$, she always wins bundle $\{1,2\}$ under $b''_1$, and obtains the same utility as under $b_1$. 
This is because bidder 1's second atomic bid $(\{1,2\}, \bar{\beta})$ is high enough to be guaranteed to win against either of the other bidders' bids, and can thus only be displaced by her own bid on bundle $\{1\}$.
But in the case when bidder 1 wins bundle $\{1\}$ under $b'_1$, she also wins bundle $\{1\}$ under $b''_1$ (due to the tiebreaking rule favoring more winners over fewer winners) and obtains strictly higher utility than with $b_1$.
The latter case happens with strictly positive probability, which proves the claim.
\Halmos
\endproof
\medskip

Connecting the proof of Theorem~\ref{thm:familyFP} back to our insights from Section~\ref{sec:complexbiddingintuition}, it becomes clear that the reason complex bidding is favored in the \family class is that the situation where bidder 1 wins her good by competing alone against bidder 3 requires a fundamentally different approach than when bidder 1 wins together with bidder 2, and only a complex bid can simultaneously optimize for these two scenarios.

\para{Introduce VCGN}
We next introduce an analogue of Theorem~\ref{thm:familyFP}, but for the VCG-nearest payment rule.
For this, we define a second class of auction domains that is almost identical to \family, except that it adds one additional constraint to bidders' valuations.

\para{Auction class VCGN (formal)}
\begin{definition}
An auction domain belongs to the \familyVCGN class if it belongs to the \family class, and additionally, bidder 2's value for $\{2\}$ is strictly smaller than 1, i.e.,
        \begin{equation}
            \label{complexfamily-vcgn}
            v_2(\{2\}) < 1.
        \end{equation}
\end{definition}

For this class, the equivalent of Theorem~\ref{thm:familyFP} also holds, when we additionally assume that the global bidder bids $b_3 = (\{1,2\}, 1)$. 
This is a very minimal and reasonable assumption, as it is known that truthful bidding is a dominant strategy for the global bidder \citep{beck2013incentives}.
%

%

\begin{theorem}
    In any auction domain of the \familyVCGN class, when using the VCG-nearest payment rule, every BNE $s^*$ where the global bidder bids truthfully contains at least one strategy $s_i^*$ that is not simple.
    \label{thm:familyVCGN}
\end{theorem}

\begin{figure}
\FIGURE
{
\begin{subfigure}[t]{0.45\textwidth}
\includegraphics[width=\textwidth]{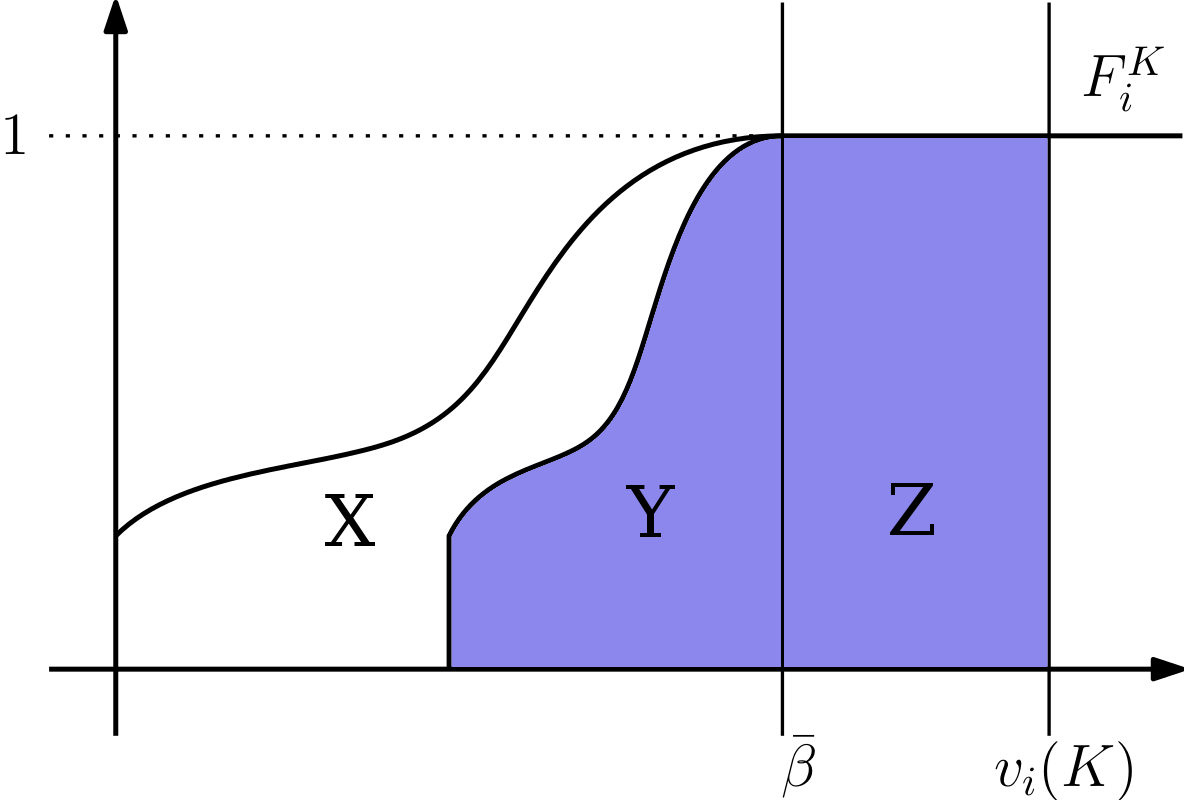}
\caption{The optimal simple bid captures all the area under the curve to the right of $\bar{\beta}$, and exactly half the area to the left of $\bar{\beta}$.}
\label{fig:case2simple}
\end{subfigure}
\hspace{3em}
\begin{subfigure}[t]{0.45\textwidth}
\includegraphics[width=\textwidth]{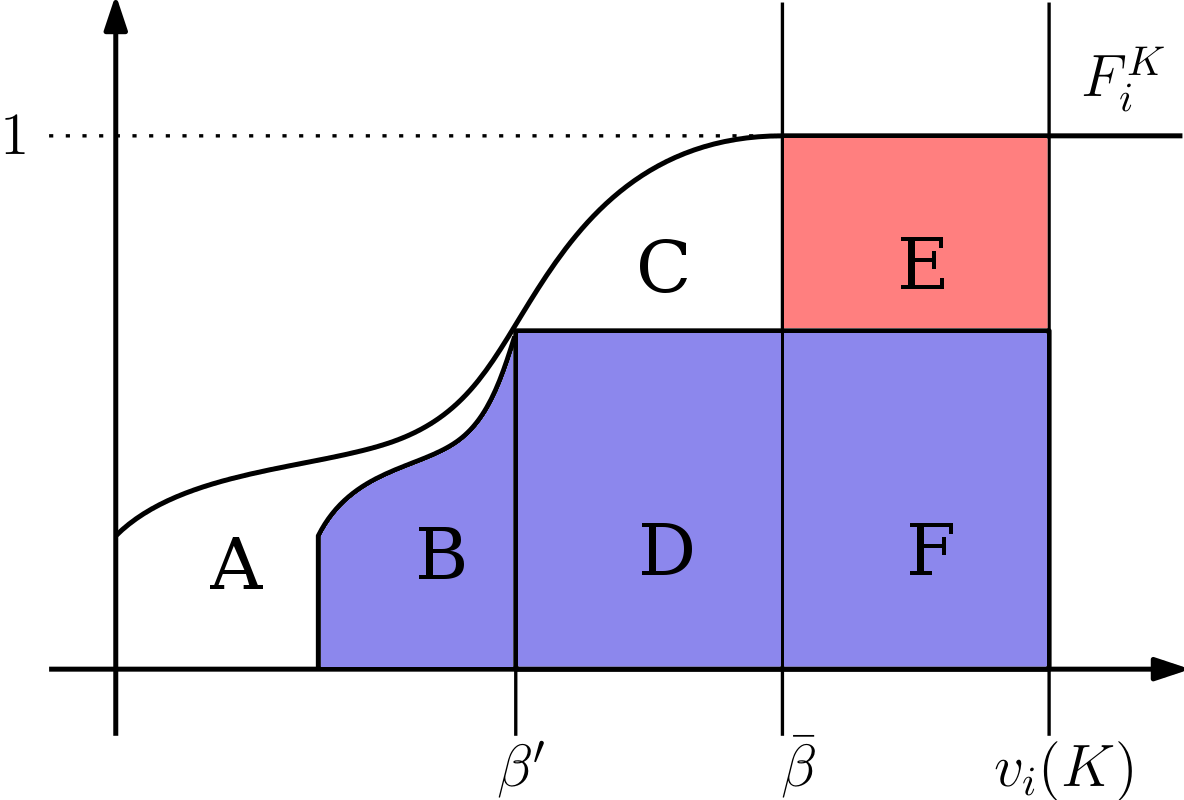}
\caption{The optimal complex bid also captures all the area to the right of $\bar{\beta}$, and strictly more than half the area to the left of $\bar{\beta}$.}
\label{fig:case2complex}
\end{subfigure}
}
{
Utility obtained by bidder 1 under the VCG-nearest payment rule, in case 2 in the proof of Theorem~\ref{thm:familyVCGN},  when the optimal simple bid is above the global bidder's threshold $\bar{\beta}$.
\label{fig:case2}
}
{}

\end{figure}

To keep the notation consistent, we still denote the global bidder's bid as $\bar{\beta}$, even though we know that $\bar{\beta} = 1$.
To prove the theorem, we must first characterize the VCG-nearest payments in the \familyVCGN class.
We do this in the electronic companion (Section~\ref{app:familyVCGN}), where we show that for a large set of bid profiles (including all those that occur in the proof of Theorem~\ref{thm:familyVCGN}), the VCG-nearest payments are exactly halfway between VCG and first price, with the first price portion being capped by $\bar\beta$:
\begin{equation}
\vcgn_i = \frac{1}{2} \cdot \left(\vcg_i + \min(b_i(\{i\}), \bar{\beta}))   \right).
\end{equation}
Geometrically, this means that, for a given bid, bidder $i$ obtains an expected utility corresponding to the whole first price rectangle below $F_i^K$, plus exactly half the area to the left of that rectangle (Figure~\ref{fig:case2}).
This fact makes it less surprising that VCG-nearest does not prevent incentives for complex bidding from arising, because the payments produced by this rule require the bidders to submit shaded bids to avoid overpaying, just like for first price.
With this in mind, we can now proceed to the proof.


\proof{Proof of Theorem~\ref{thm:familyVCGN}.}
\para{VCGN is analogous to FP}
The proof proceeds analogously to that of Theorem~\ref{thm:familyFP}.
When we derived the constraints on the shape of $F_1^{\{1\}}$ and $F_1^{\{1,2\}}$, it required us to argue first that $b_2(\{2\}) \leq \bar{\beta}$, and then that $b_2(\{2\}) \leq v_2(\{2\})$.
Under first price, this was very straightforward. 
We now show the same for VCG-nearest as follows: the first inequality follows from the second, due to our new assumption (\ref{complexfamily-vcgn}).
For the second inequality, we combine (\ref{complexfamily-vcgn}) and Lemma~\ref{lem:vcgn_simple} to show that the payment strictly increases when going from the truthful bid to a higher one, and thus the argument we made for first price also applies.

Next, we again consider $b_1 = s_1(v_1) = (\{1\}, \beta)$, and split the remainder of the proof into two cases.

\textbf{Case 1} ($\beta < \bar{\beta}$):
This case proceeds exactly like in the proof of Theorem~\ref{thm:familyFP}, except that we need to make use of the just established fact that $b_2(\{2\}) \leq \bar{\beta}$ to show that $q_4 = 0$.
We then use Lemma~\ref{lem:vcgn_complex} to determine the payments and thus show that $b'_1 = (\{1\}, \beta + \epsilon) \oplus (\{1, 2\}, \bar{\beta} + \epsilon)$ obtains more expected utility than $b_1$ in the limit $\epsilon \to 0$.

\todo{NOTE: Currently omitting detailed derivation, because it has not been updated. Should make the expressions for the payments exact, but that makes it harder to extract the $\epsilon$.}

\textbf{Case 2} ($\beta \geq \bar{\beta}$):
We show geometrically that there exists a complex bid $b''_1 = (\{1\}, \beta') \oplus (\{1,2\}, \bar{\beta})$ for some $0 < \beta' <  \bar{\beta}$ that captures strictly more utility than $b_1 = (\{1\}, \beta)$.
The utilities of $b_1$ and $b_1''$ are illustrated in Figure~\ref{fig:case2}, where we used Lemmas~\ref{lem:vcgn_simple} and~\ref{lem:vcgn_complex} to determine the payments.

As observed before, bidder 1 captures the entire area under the curve to the right of $\bar{\beta}$, plus half the area to the left of $\bar{\beta}$ under the optimal simple bid $b_i$.
This splits the area under $F_1^{\{1\}}$ into three pieces $X, Y$ and $Z$, and the utility captured is $|Y| + |Z|$ (Figure~\ref{fig:case2simple}).

For the complex bid $b_1''$, we observe the following:
when winning with the atomic bid $(\{1\}, \beta')$ bidder 1 captures the entire rectangle to the right of $\beta'$ and below $F_1^{\{1\}}(\beta')$, as well as half the area to the left of $\beta'$.
When winning with the global bundle $\{1,2\}$, the payment is $\bar{\beta}$, so bidder 1 additionally captures the rectangle to the right of $\bar{\beta}$ and above $F_1^{\{1\}}(\beta')$.
This splits the area under $F_1^{\{1\}}$ into six pieces $A, B, C, D, E$ and $F$, and the utility captured is  $|B| + |D| + |E| + |F|$ (Figure~\ref{fig:case2complex}).

If we want $b_1''$ to yield strictly higher utility than $b_1$, we need to ensure that
\begin{equation}
    |B| + |D| > |Y|,
    \label{eq:areasy1}
\end{equation}
since we have the obvious identity $|E| + |F| = |Z|$.
Now, it is clear that
\begin{equation}
    |X| + |Y| = |A| + |B| + |C| + |D|,
    \label{eq:areasy1.5}
\end{equation}
and Lemma~\ref{lem:vcgn_simple} tells us that the VCG-nearest payments are exactly between VCG and first price, so $|A| = |B|$ and $|X| = |Y|$.
Thus, \eqref{eq:areasy1.5} can be rewritten to
\begin{equation}
    2|Y| = 2|B| + |C| + |D|,
    \label{eq:areasy2}
\end{equation}
which can be combined with \eqref{eq:areasy1} and simplified to
\begin{equation}
    |D| > |C|.
    \label{eq:areasy3}
\end{equation}

The only thing still to be shown is that there always exists $\beta'$ such that \eqref{eq:areasy3} holds.
Let $s$ be the supremum of $F_1^{\{1\}}$ in the open interval $[0, \bar{\beta})$.
By definition of a supremum, there must exist $\beta'$ such that $F_1^{\{1\}}(\beta') > s/2$.
But this means that the area between $\beta'$ and $\bar{\beta}$ is at most $s \cdot (\bar{\beta} - \beta')$, and $|D| = F_1^{\{1\}}(\beta') \cdot (\bar{\beta} - \beta')$ is strictly more than half of that.
\Halmos
\endproof

\medskip


Theorem~\ref{thm:familyVCGN} provides a strong headwind for the conventional wisdom that VCG-nearest makes bidding in a CA particularly easy.
Furthermore, we observe that the same logic underlying the incentives for complex bidding also applies in situations where all other bidders simply report truthfully:

\begin{proposition}
    In an auction domain of the \familyVCGN class, under the first price or VCG-nearest payment rules, bidder 1's best response against the truthful strategy profile $s_\smi(V_\smi) = V_\smi$ is not simple.
    \label{prop:aeou}
\end{proposition}
This result immediately follows by the same arguments of the proofs of Theorems~\ref{thm:familyFP} and~\ref{thm:familyVCGN}, and we therefore omit a formal proof.

\section{Exponential Separation of Utilities with Simple vs.\ Complex Bids}

\label{sec:mainresult}


So far, we have established that complex bidding is systematically incentivized in many auction instances.
We now ask a natural follow-up question: can we put a bound on how large this effect is in the worst case?
In Section~\ref{sec:complexbiddingintuition}, we showed that a bidder can exploit the correlations between winning thresholds induced by the winner determination problem as a signal to activate the best among two conditional bids.
At first sight, we might hope that there is a tight upper bound on how much utility can be gained via complex bidding -- maybe a factor two, given that the signal we used was only \emph{binary}.
However, as we show next, it is possible for a signal to encode many more than just two different world states, thus leading to a much larger increase in utility.

We now return to the point of view we adopted in Section~\ref{sec:thresholds}, considering a single bidder $i$'s decision problem when faced with a fixed distribution of bids $B_\smi$.
This helps us better understand just how large a complex bid can get (in terms of the number of XOR atoms), and still have each of those atoms be relevant for the total utility obtained by the bidder.
For this, we construct a family of auction instances (i.e., a series of auctions, parameterized by $m$, the number of goods being sold).
Surprisingly, we show that the size of the optimal complex bid can be exponential in $m$, and the complex best response utility is exponentially larger than the simple best response utility.

\subsection{First Price}

We first state this result for the first price payment rule:

\begin{theorem}
    There exists an auction family $F(m)$ with $m$ goods and using the first price payment rule, such that for some single-minded bidder $i$ and bid distribution $B_\smi$, we have that
    \begin{equation}
        \frac{\eu_i^c(v_i,B_\smi)}{\eu_i^s(v_i,B_\smi)} = \Theta \left(2^m\right).
    \end{equation}
    Furthermore, the complex best response must contain $\Theta(2^m)$ times more atoms than the simple best response.
    \label{thm:FP}
\end{theorem}

\begin{figure}
\FIGURE
{
\begin{subfigure}[t]{0.29\textwidth}
\centering
\includegraphics[width=\textwidth]{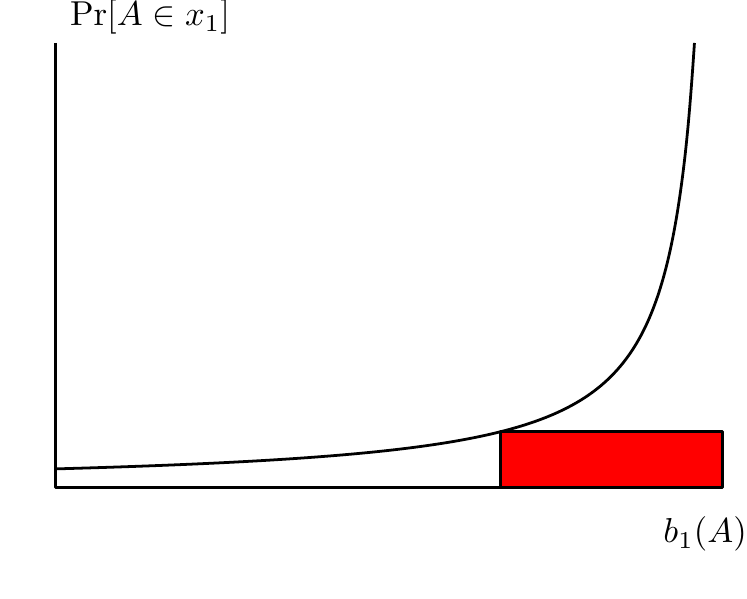}
\caption{A low bid obtains a high payoff with low probability.}
\end{subfigure}
\qquad
\begin{subfigure}[t]{0.29\textwidth}
\centering
\includegraphics[width=\textwidth]{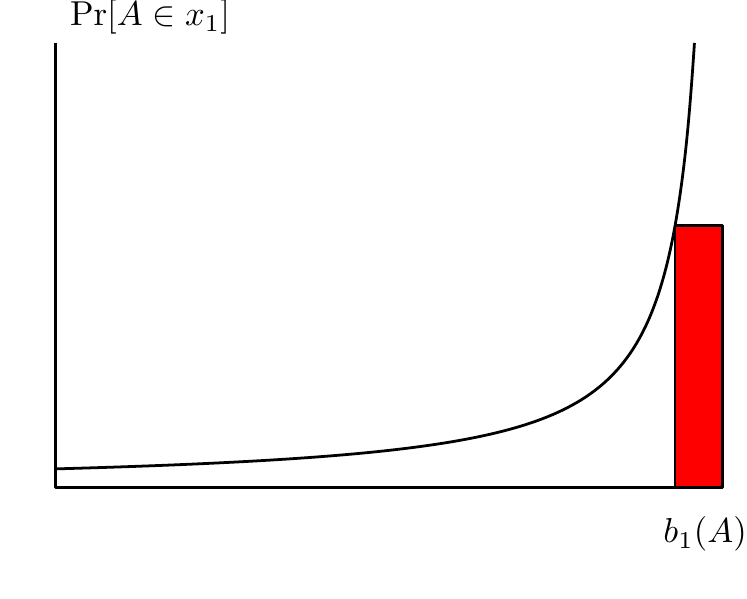}
\caption{A high bid obtains a low payoff with high probability.}
\end{subfigure}
\qquad
\begin{subfigure}[t]{0.29\textwidth}
\centering
\includegraphics[width=\textwidth]{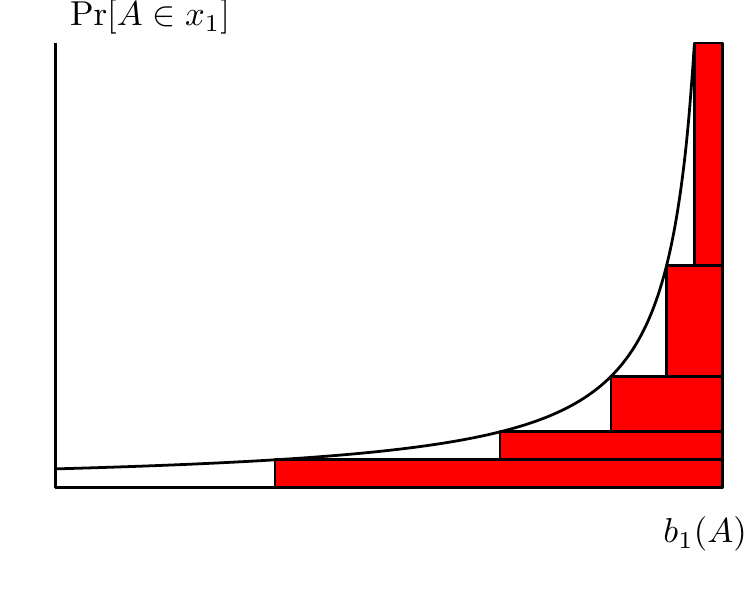}
\caption{A complex bid obtains the highest possible payoff.}
\end{subfigure}
}
{
Construction of the distribution $F_1^K$ used in Theorem~\ref{thm:FP}.
The area under the curve is shaped such that no simple bid can capture a large fraction of the total utility, but a complex bid with exponentially many atomic bids can capture the whole area.
\label{fig:construction}
}
{}

\end{figure}

Before we proceed to the proof of Theorem~\ref{thm:FP}, we provide an overview of the main proof ideas.
The full proof can be found below in Section~\ref{sec:FP}.
We construct an auction instance where the bid profile $B_{\sm1}$ encodes a signal that reveals the exact winning threshold $\vcg_1(B_{\sm 1}, \{m+1\})$ to bidder 1.
Given this signal, bidder 1 can make a separate atomic bid for each possible realization of the winning threshold, and pack them into one XOR bid in such a way that the atomic bids do not interfere with each other, i.e., each of them wins if and only if its associated threshold is realized.

To maximize the ratio of the simple and complex best response utilities, we lay out the realizations of the winning thresholds along a geometric series.
See Figure~\ref{fig:construction} for a graphical representation of our construction.
To minimize the utility of the simple best response, we balance $F_1^{\{m+1\}}$ such that the payoff obtained by a bid is inversely proportional to the winning probability.
Consequently, all simple bids yield the same (low) expected utility, regardless of the bid amount.
Conversely, to maximize the complex best response utility, we ensure that there are exponentially many different winning thresholds, and that the utility rectangles associated to each of them overlap very little with each other.
Thus, each atomic bid in the complex best response obtains almost as much utility as the simple best response by itself.

\subsection{Proof of Theorem~\ref{thm:FP} (First Price)}
\label{sec:FP}



We introduce an auction family, i.e., a single-parameter series of instances $F(m+1)$ of increasing size.\footnote{We define the family in terms of the number of goods $m+1$ (for $m \in \mathbb{N}$) only to simplify the notation.
The resulting offset of 1 can be folded into the constant factor implicit in the Landau notation $\Theta$.}
%
%
Let $m \in \mathbb{N}$. For $i \in \mathbb{R}_+$, let $f(i) := 1 - \frac{1}{2^i}$, let $C$ be a large constant, and let $\mathcal{K}$ be the powerset of the first $m$ goods, i.e., the set of all bundles $K \subseteq \{1,\ldots, m\}$ which do not contain good $m+1$.
Then, let $\sigma$ be a bijective function such that
\begin{align}
& \sigma: \mathcal{K} \mapsto \{1, \ldots,2^m\}\\
& K' \subset K \Rightarrow \sigma(K') > \sigma(K).
\end{align}
In other words, $\sigma$ assigns a rank to each bundle in $\mathcal{K}$, such that this ranking forms a total order consistent with the partial order induced by the subset relation.

\begin{definition}
The auction family $F(m+1)$ is given as follows:
\begin{enumerate}
\item The auction has $3$ bidders and $m + 1$ goods.
\item Bidder 1 has value $1$ for bundle $\{m+1\}$, and no interest in any other bundle.
\item Bidder 2 always bids $C$ on the bundle of all goods.
\item Bidder 3 bids $C - f(j)$ on bundle $\sigma^{-1}(j)$, where $j$ is a random variable chosen from $\{1,\ldots,2^m\}$ with probability proportional to $2^j$.
This can be achieved by normalizing $2^j$ with the factor $\sum_{j=1}^{2^m} 2^j = 2^{2^m+1}-2$.

\item The allocation rule is efficient, and the payment rule is first-price.
\end{enumerate}
\label{def:fp-family}
\end{definition}


\begin{table}
\TABLE
{
Bundle bids made by different bidder types in auction family $F(m+1)$.
Bidder 1 works together with bidder 3 (the helper). If bidder 1 bids $f(i)$, they can jointly defeat the competitor whenever $i \geq j$. Bidder 1 can make additional bundle bids that include more goods, and those bids are active whenever they do not overlap with the helper's bid.
\label{tab:fp}
}
{
\begin{tabular}{lc||ccccc}

\multicolumn{2}{c||}{\bf Bidders} &
\multicolumn{2}{c}{\bf Goods} \\

&&
\bf Signalling Goods &
\bf Prize \\

&&
$1$
\ldots
$m$ &
$m+1$ \\
\hhline{=:=::=:=}

\bf Protagonist &1 & \multicolumn{1}{c|}{} & \multicolumn{1}{c|}{$f(i)$}\\
\cline{3-4}

\bf Competitor &2 & \multicolumn{2}{c|}{$C$} \\
\cline{3-4}

\bf Helper  & 3 & \multicolumn{1}{c|}{$C - f(j)$ on $\sigma^{-1}(j)$} & \multicolumn{1}{c}{}\\
\cline{3-3}

\end{tabular}
}
{}
\end{table}

The bidders of this auction family and the relation of their bundles to each other are shown in Table~\ref{tab:fp}.
To understand the asymptotic behaviour of this auction family, we first need to understand the strategic landscape that bidder 1 is presented with. For this, we must formally define which bids even have a chance of winning.
For our next definition, recall that $W(b, L, K)$ is the reported social welfare achieved when coalition $L$ jointly wins the goods $K$.

\begin{definition}
    Given a bid profile $b$, a bundle $K$ is \emph{active} for bidder 1 if $m+1 \in K$ and
    \begin{align}
        W(b, N \setminus \{1\}, M \setminus K) = W(b, N \setminus \{1\}, M \setminus \{m+1\}).
    \end{align}
    \label{def:active}
\end{definition}
In words, a bundle is active whenever it does not matter (for winner determination), whether bidder 1 makes an atomic bid on her original bundle of interest $\{m+1\}$ or on bundle $K$.
In the latter case, she would just win some unneeded goods on top of good $m+1$.
It follows that bidder 1 can combine any bid on an active bundle with the bid by bidder 3, to jointly overcome the threshold needed to win against bidder 2.
Given this definition, we now characterize the auction's possible outcomes in the following lemma.

\begin{lemma}
    In the auction family $F(m+1)$, let $b_1$ be a bid for bidder 1, and let $K^*$ be the active bundle with the highest atomic bid in $b_1$.
    Furthermore, assume that bidder 1's bids on other bundles are not too high, i.e., $\forall K \ne K^*: b_1(K) \leq 1$.
    
    Then, bidder 1 wins bundle $K^*$ iff $b_1(K^*) \geq f(j)$.
    Otherwise, bidder 1's allocated bundle does not include good $m+1$.
    \label{lem:fp-payments}
\end{lemma}

\proof{Proof.}
    Because bidder 1's bids are capped at $1$, she cannot win without the help of bidder 3.
    It is always a feasible allocation to give all goods to bidder 2, for a reported social welfare of $C$.
    This welfare is exceeded by a combination of bidder 1 and 3's bids exactly when the bid on the largest active bundle $K^*$ is at least $f(i)$ for some $i \geq j$.
    \Halmos
\endproof

Note that due to the first-price payment rule, bidder 1's payment does not depend on the identity of $K^*$, but only on the bid amount $b_1(K^*)$.
Furthermore, if $b_1$ contains any atomic bid larger than $1$, bidder 1 would either never win with this bid, making it irrelevant, or she would win with negative utility.
Thus, we can assume WLOG that $b_1$ fulfills the assumption of Lemma~\ref{lem:fp-payments}, and we can easily compute the utility obtained by bidder 1 for any given realization of $j$.
With this information, we can now establish bidder 1's simple and complex best response utilities.

\begin{lemma}
    In auction family $F(m+1)$, bidder 1's simple best response utility is $\displaystyle \Theta\left(\frac{1}{2^{2^m}}\right)$.
    \label{lem:fp_singleminded_brutil}
    \vspace{2mm}
\end{lemma}

\proof{Proof.}
    Bidder 1's payment increases monotonically in $i$, so it is optimal for her to bid exactly on one of the levels $f(i)$ for $i \in \mathbb{N}$, where the winning probability makes a discrete jump.
    Note that any bid between two bid levels (or above level $f(2^m)$) is strictly worse than the next lower bid level.

    For a bid of $f(i)$, bidder 1 wins whenever $i \geq j$ and has a utility of $1 - (1 - \frac{1}{2^i}) = \frac{1}{2^i}$ (because the payment rule is first-price and utilities are quasi-linear).
    Recall that the probability distribution of $j$ is proportional to $2^j$.
    Thus, bidder 1's expected utility for any  $1 \leq i \leq 2^m$ is
    \begin{equation}
        \sum_{j=1}^i \frac{2^j}{2^{2^m+1} -2} \cdot \frac{1}{2^i}
        \enskip=\enskip
        \frac{1}{2^{2^m+1} -2} \cdot \frac{1}{2^i} \sum_{j=1}^i 2^j
        \enskip=\enskip
        \Theta\left( \frac{1}{2^{2^m}} \right).
    \end{equation}
    \Halmos
\endproof

\begin{lemma}
    In auction family $F(m+1)$, bidder 1's best response utility is $\displaystyle \Theta\left(\frac{2^m}{2^{2^m}}\right)$.
    \label{lem:fp_full_brutil}
    \vspace{2mm}
\end{lemma}

\proof{Proof.}
    Since bidder 1's payment is independent of the winning bundle's identity, her utility is maximized when she bids in such a way that for each $j$, she wins with an atomic bid that offers just the minimum amount needed to win.
    This bound can be achieved by bidding as follows:
    For each bundle $K$ that contains good $m+1$ (i.e., all bundles \emph{not} contained in $\mathcal{K}$), bidder 1 adds an atomic bid $(K, f(\sigma(M\setminus K)))$ to her bid.
    This bid is just high enough to win when the signal is $j = \sigma(M \setminus K)$.
    Furthermore, this atomic bid does not win with any other realization of the signal.
    
    We show this by case split over the realizations of $j$.
    Case 1: We have that $j = \sigma(M \setminus K')$ for some $K' \not \supset K$. 
    Since $K' \neq K$, there exists at least one good that is in $K$ but not in $K'$. Any bundle bid on $K$ cannot win because it is not active (it overlaps with bidder 3's bid on $M\setminus K'$), and would need to be at least $C$.
    Case 2: We have that $j = \sigma(M \setminus K')$ for some $K' \supset K$.
    The bundle bid on $K$ is dominated (in the winner determination problem) by the bundle bid on $K'$, 
    because (due to $\sigma$ respecting the partial order of $\mathcal{K}$ and $f$ being monotonic), we have that
    \begin{equation}
    K' \supset K \Rightarrow M \setminus K' \subset M \setminus K \Rightarrow \sigma(M \setminus K') > \sigma(M \setminus K) \Rightarrow f(\sigma(M \setminus K')) > f(\sigma(M \setminus K)).
    \end{equation}
    It follows that for each realization of $j$, the highest active bid is $f(j)$.
    Applying Lemma~\ref{lem:fp-payments}, the expected utility is thus
    \begin{equation}
        \sum_{j=1}^{2^m} \frac{2^j}{2^{2^m+1} -2} \cdot \frac{1}{2^j}
        \enskip=\enskip 
        \frac{1}{2^{2^m+1} -2} \cdot \sum_{j=1}^{2^m} 1 
        \enskip=\enskip
        \Theta\left( \frac{2^m}{2^{2^m}} \right).
    \end{equation}
    \Halmos
\endproof

With these lemmas in place, we can now prove the exponential separation for the first price payment rule.

\proof{Proof of Theorem~\ref{thm:FP}.}
    Consider the family of auctions $F(m+1)$.
    The total number of goods is $m + 1 = \Theta(m)$.
    Using Lemmas~\ref{lem:fp_singleminded_brutil} and~\ref{lem:fp_full_brutil}, the ratio of the simple and complex best response utilities of bidder 1 is
    \begin{equation}
           \Theta\left( \frac{2^m}{2^{2^m}} \middle/ \frac{1}{2^{2^m}}  \right)
        = \Theta\left( 2^m \right).
    \end{equation}
    It follows from Lemma~\ref{lem:fp_singleminded_brutil} by union bound that $\Theta(2^m)$ atoms are needed to achieve the complex best response utility.
%
    \Halmos
\endproof

\subsection{VCG-nearest}

It turns out that, as in Section~\ref{sec:complexfamily}, the equivalent result for VCG-nearest also holds:

\begin{theorem}
    There exists an auction family $F'(m)$ with $m$ goods and using the VCG-nearest payment rule, such that for some single-minded bidder $i$ and bid distribution $B_\smi$, we have that
    \begin{equation}
        \frac{\eu_i^c(v_i,B_\smi)}{\eu_i^s(v_i,B_\smi)} = \Theta \left(2^m\right).
    \end{equation}
    Furthermore, the complex best response must contain $\Theta(2^m)$ times more atoms than the simple best response.
    \label{thm:VCG-nearest}
\end{theorem}

The proof for Theorem~\ref{thm:VCG-nearest} proceeds along the same lines as for Theorem~\ref{thm:FP}, except that we need to construct the auction family in such a way that the payment rule behaves very badly for bidder 1.
Showing this requires some additional technical work, such as a detailed analysis of all the core constraints involved, and we provide the proof in the electronic companion (Section~\ref{app:VCG-nearest-proof}). 
\smallskip


Theorems~\ref{thm:FP} and~\ref{thm:VCG-nearest} are striking results because they demonstrate that even a single-minded bidder can make productive use of the entire exponentially large strategy space offered by CAs.
Thus, we cannot provide an upper bound on the profitability of complex bidding manipulations better than $2^m$.\footnote{Even this bound only holds for the first price payment rule, where the payments are always independent of losing bids. For VCG-nearest, we cannot even guarantee this seemingly trivial bound, since the payment of a bidder may be affected by her own atomic bids on other bundles, leaving the door open for even more extreme manipulations.}
It is still possible that a tighter bound (e.g., a constant factor difference between the simple and complex best response utility) could be established in BNE.
However, any result in this direction will require an explicit analysis of the equilibrium conditions, as the single bidder perspective we take does not impose strong enough constraints on the problem.


\section{Discussion}
\label{sec:discussion}

Our results from Sections~\ref{sec:complexfamily} and~\ref{sec:mainresult} are structural in essence: they are mostly derived from the combinatorial properties of the winner determination problem, and the way that multiple atomic bids from the same bidder compete with each other, based on their relative strength compared to the winning thresholds of their respective bundles.
This is a fundamental property of CAs, independent of more superficial aspects of the problem, and thus very difficult to avoid.

For convenience and clarity, we have provided our definition of simple and complex bidding with respect to the XOR bidding language.
However, it is possible to define complex bidding only in terms of the marginal bids implied by whatever bidding language is used to report the valuations to the auctioneer.
In Section~\ref{sec:mainresult}, our results regarding the size of a bid (in terms of number of atomic bids) do depend on the bidding language, but can easily be adapted to other languages such as OR*.
The remainder of our results only require that the bidding language be fully expressive, able to represent all possible valuations.

Our results in Section~\ref{sec:complexfamily} are ``almost tight'': for the \familyVCGN class, conditions \eqref{complexfamily-val1} and \eqref{complexfamily-v2_not_always_zero} are \emph{required} for bidder 1 to have any incentives for complex bidding under the VCG-nearest payment rule. In the absence of either condition, bidder 1's strategy space collapses to one dimension.
For \family and first price, the same idea holds in principle, with some technical caveats.
Thus, the \family and \familyVCGN classes cannot be made substantially larger, as they already cover most situations where a bidder has incentives for complex bidding in LLG-like domains.
The fact that the global bidder has a fixed valuation is also not essential, and only made to simplify our analysis and make our findings cleaner.

To ground the significance of our results, we point out that LLG-like dynamics often arise naturally in real-world auctions.
For instance, \citet{ausubel2020vcg} provide examples of zero-revenue outcomes that have occured in spectrum auctions, where the bidders exhibit valuations closely mimicking LLG-like settings in the assignment stage of the combinatorial clock auction.

Regarding our results from Section~\ref{sec:mainresult}, we note that Theorems~\ref{thm:FP} and~\ref{thm:VCG-nearest} immediately translate to deviations from \emph{truthful} bidding as well, by simply setting the valuations of bidders 2 and 3 such that the bid profile used in the proof is truthful. 
This contradicts the argument found in the literature that straightforward truthful bidding is a reasonable strategy under this rule \citep{Cramton2013SpectrumAuctionDesign, Milgrom2007PackageAuctionsAndExchanges}.

The equivalent of Theorem~\ref{thm:VCG-nearest} might also hold for other minimum-revenue core-selecting payment rules.
Many of the arguments in the proof are not specific to VCG-nearest but simply establish that the shape and location of the core can be forced to be very asymmetric between bidders.
Thus, any payment rule that computes payments by projecting a reference point to the core ``from below'' is potentially vulnerable to the same construction.
However, some minimum-revenue core-selecting payment rules such as nearest-bid \citep{ausubel2020core} project to the minimum-revenue core ``from above,'' and might thus be less susceptible to such asymmetric constructions, creating fewer incentives for complex bidding.

\medskip
\section{Conclusion}

\todo{Future work: More precise definition of complex bidding vs overbidding.}

In this paper, we have analyzed the existence and magnitude of complex bidding manipulations in CAs.
These manipulations are highly counter-intuitive, but they are incentivized, because they enable bidders to circumvent the uncertainty they face regarding the bids of others.
We have presented two pieces of evidence that complex bidding can be a major issue:
first, there exist entire classes of auction domains where simple bidding is dominated by complex bidding in BNE.
Second, the large strategy space available to bidders in a CA allows them a lot of room to manipulate via complex bids, leading to an exponential utility increase in the worst case.

It is particularly surprising that VCG-nearest does not provide any protection against this manipulation, given that it is a minimum-revenue core-selecting payment rule.
In particular, we have shown that, in the domains we have studied, VCG-nearest can be forced to behave identically to first price for some of the bidders, losing its conceptual similarity to second price payment rules.
It is an interesting topic for future work to study whether other core-selecting payment rules are much better behaved in this sense or whether this is a fundamental property of all such rules.

Our findings raise many questions for practical auction design, given that the strategic behavior we describe could interfere with efficiency and price discovery, and might give a major advantage to the most sophisticated bidders.
Taking a broader mechanism design point of view, we note that \emph{any} mechanism that produces efficient outcomes (even if only in BNE) must implicitly or explicitly solve the winner determination problem, and must produce the signals that our manipulation depends on.
It is thus natural to ask in what kind of form these signals are represented in different mechanisms, and whether there are principled ways how this information can be isolated and made inaccessible to bidders.

While our exponential separation of utility is an extreme example that we do not expect to encounter as-is in practice, the logical pathway underpinning it is quite robust.
Therefore, we expect that opportunities for complex bidding manipulations also arise in real-world CAs, especially in settings with heterogeneous bidders and sufficiently rich valuations.
However, many practical applications of auctions use domain-specific bidding languages, or impose other restrictions on bidders' strategies.
For example, spectrum auctions typically have a bound on the size of bids on the order of 500 atomic bids \citep{ausubel2017practical}.
Under these types of constraints, it might be possible to establish bounds on how profitable complex bidding manipulations can be.
Similarly, the benefit of complex bidding could be bounded in terms of the structure and complexity of bidders' value distributions that arise in different application domains.
Taking such considerations into account, we believe it might be possible to design mechanisms and bidding languages that make it hard to report valuations that are unnatural or overly complex, while still allowing for enough expressivity for bidders to accurately report their true valuations.
Succeeding at this difficult task would preserve the known advantages of combinatorial auctions, while restoring the natural desideratum of simplicity.





\bibliographystyle{informs2014} 

\bibliography{CApapers}

\begin{thebibliography}{41}
\providecommand{\natexlab}[1]{#1}
\providecommand{\url}[1]{\texttt{#1}}
\providecommand{\urlprefix}{URL }

\bibitem[{Ausubel \protect\BIBand{} Baranov(2017)}]{ausubel2017practical}
Ausubel L, Baranov O (2017) A practical guide to the combinatorial clock
  auction. \emph{The Economic Journal} 127(605).

\bibitem[{Ausubel et~al.(2006)Ausubel, Cramton, \protect\BIBand{}
  Milgrom}]{ausubel2006clock}
Ausubel L, Cramton P, Milgrom P (2006) The clock-proxy auction: A practical
  combinatorial auction design. Cramton P, Shoham Y, Steinberg R, eds.,
  \emph{Combinatorial Auctions} (MIT Press).

\bibitem[{Ausubel et~al.(2011)Ausubel, Cramton et~al.}]{ausubel2011auction}
Ausubel L, Cramton P, et~al. (2011) Auction design for wind rights.
  \emph{Report to Bureau of Ocean Energy Management, Regulation and
  Enforcement} .

\bibitem[{Ausubel \protect\BIBand{} Milgrom(2006)}]{ausubel2006lovely}
Ausubel L, Milgrom P (2006) The lovely but lonely {V}ickrey auction. Cramton P,
  Shoham Y, Steinberg R, eds., \emph{Combinatorial Auctions} (MIT Press).

\bibitem[{Ausubel \protect\BIBand{}
  Baranov(2020{\natexlab{a}})}]{ausubel2020core}
Ausubel LM, Baranov O (2020{\natexlab{a}}) Core-selecting auctions with
  incomplete information. \emph{International Journal of Game Theory}
  49(1):251--273.

\bibitem[{Ausubel \protect\BIBand{}
  Baranov(2020{\natexlab{b}})}]{ausubel2020vcg}
Ausubel LM, Baranov O (2020{\natexlab{b}}) Vcg, the core, and assignment stages
  in auctions Working Paper.

\bibitem[{Babaioff et~al.(2021)Babaioff, Gonczarowski, \protect\BIBand{}
  Nisan}]{babaioff2021menu}
Babaioff M, Gonczarowski YA, Nisan N (2021) The menu-size complexity of revenue
  approximation. \emph{Games and Economic Behavior} .

\bibitem[{Babaioff et~al.(2020)Babaioff, Immorlica, Lucier, \protect\BIBand{}
  Weinberg}]{babaioff2020simple}
Babaioff M, Immorlica N, Lucier B, Weinberg SM (2020) A simple and
  approximately optimal mechanism for an additive buyer. \emph{Journal of the
  ACM (JACM)} 67(4):1--40.

\bibitem[{Baranov(2010)}]{baranov2010exposure}
Baranov O (2010) Exposure vs. free-riding in auctions with incomplete
  information, working paper.

\bibitem[{Beck \protect\BIBand{} Ott(2013)}]{beck2013incentives}
Beck M, Ott M (2013) Incentives for overbidding in minimum-revenue
  core-selecting auctions. \emph{Annual Conference 2013 (Duesseldorf):
  Competition Policy and Regulation in a Global Economic Order} .

\bibitem[{Bichler et~al.(2021)Bichler, Fichtl, Heidekr{\"u}ger, Kohring,
  \protect\BIBand{} Sutterer}]{bichler2021learning}
Bichler M, Fichtl M, Heidekr{\"u}ger S, Kohring N, Sutterer P (2021) Learning
  equilibria in symmetric auction games using artificial neural networks.
  \emph{Nature Machine Intelligence} .

\bibitem[{B{\"o}rgers \protect\BIBand{} Li(2019)}]{borgers2019strategically}
B{\"o}rgers T, Li J (2019) Strategically simple mechanisms. \emph{Econometrica}
  87(6):2003--2035.

\bibitem[{Bosshard et~al.(2020)Bosshard, B\"{u}nz, Lubin, \protect\BIBand{}
  Seuken}]{Bosshard2020JAIR}
Bosshard V, B\"{u}nz B, Lubin B, Seuken S (2020) Computing bayes-nash
  equilibria in combinatorial auctions with verification. \emph{Journal of
  Artificial Intelligence Research} 69:531--570.

\bibitem[{Bosshard \protect\BIBand{} Seuken(2021)}]{Bosshard2021CostEC}
Bosshard V, Seuken S (2021) The cost of simple bidding in combinatorial
  auctions. \emph{Proceedings of the 22nd ACM Conference on Electronic Commerce
  (EC)}.

\bibitem[{Bosshard et~al.(2018)Bosshard, Wang, \protect\BIBand{}
  Seuken}]{Bosshard2018nondecreasing}
Bosshard V, Wang Y, Seuken S (2018) Non-decreasing payment rules in
  combinatorial auctions. \emph{Proceedings of the 27th International Joint
  Conference on Artificial Intelligence (IJCAI)} (Stockholm, Sweden).

\bibitem[{B\"{u}nz et~al.(2018)B\"{u}nz, Lubin, \protect\BIBand{}
  Seuken}]{lubin2018designing}
B\"{u}nz B, Lubin B, Seuken S (2018) Designing core-selecting payment rules: A
  computational search approach. \emph{Proceedings of the 19th ACM Conference
  on Electronic Commerce (EC)} (Ithaca, NY, USA).

\bibitem[{Cantillon \protect\BIBand{}
  Pesendorfer(2007)}]{cantillon2007combination}
Cantillon E, Pesendorfer M (2007) Combination bidding in multi-unit auctions .

\bibitem[{Clarke(1971)}]{clarke1971multipart}
Clarke E (1971) Multipart pricing of public goods. \emph{Public Choice}
  11(1):17--33.

\bibitem[{Cramton(2013)}]{Cramton2013SpectrumAuctionDesign}
Cramton P (2013) Spectrum auction design. \emph{Review of Industrial
  Organization} 42(2):161--190.

\bibitem[{Cramton et~al.(2006)Cramton, Shoham, \protect\BIBand{}
  Steinberg}]{CramtonEtAl2006CombAuctions}
Cramton P, Shoham Y, Steinberg R (2006) \emph{Combinatorial Auctions} (MIT
  Press).

\bibitem[{Day \protect\BIBand{} Cramton(2012)}]{DayCramton2012Quadratic}
Day RW, Cramton P (2012) Quadratic core-selecting payment rules for
  combinatorial auctions. \emph{Operations Research} 60(3):588--603.

\bibitem[{Day \protect\BIBand{}
  Milgrom(2008)}]{DayMilgrom2008CoreSelectPackageAuctions}
Day RW, Milgrom P (2008) Core-selecting package auctions. \emph{International
  Journal of Game Theory} 36(3):393--407.

\bibitem[{Day \protect\BIBand{} Raghavan(2007)}]{DayRaghavan2007FairPayments}
Day RW, Raghavan S (2007) Fair payments for efficient allocations in public
  sector combinatorial auctions. \emph{Management Science} 53(9):1389--1406.

\bibitem[{Epstein et~al.(2004)Epstein, Henr{\'\i}quez, Catal{\'a}n, Weintraub,
  Mart{\'\i}nez, \protect\BIBand{} Espejo}]{epstein2004combinatorial}
Epstein R, Henr{\'\i}quez L, Catal{\'a}n J, Weintraub GY, Mart{\'\i}nez C,
  Espejo F (2004) A combinatorial auction improves school meals in chile: a
  case of or in developing countries. \emph{International Transactions in
  Operational Research} 11(6):593--612.

\bibitem[{Goeree \protect\BIBand{}
  Lien(2016)}]{Goeree2013OnTheImpossibilityOfCoreSelectingAuctions}
Goeree J, Lien Y (2016) On the impossibility of core-selecting auctions.
  \emph{Theoretical Economics} 11:41--52.

\bibitem[{Goetzendorff et~al.(2015)Goetzendorff, Bichler, Day,
  \protect\BIBand{} Shabalin}]{goetzendorff2014core}
Goetzendorff A, Bichler M, Day R, Shabalin P (2015) Compact bid languages and
  core-pricing in large multi-item auctions. \emph{Management Science}
  61(7):1684--1703.

\bibitem[{Groves(1973)}]{groves1973incentives}
Groves T (1973) Incentives in teams. \emph{Econometrica} 41(4):617--631.

\bibitem[{Janssen \protect\BIBand{} Karamychev(2016)}]{janssen2016spiteful}
Janssen M, Karamychev V (2016) Spiteful bidding and gaming in combinatorial
  clock auctions. \emph{Games and Economic Behavior} 100:186--207.

\bibitem[{Kokott et~al.(2017)Kokott, Bichler, \protect\BIBand{}
  Paulsen}]{kokott2017combinatorial}
Kokott GM, Bichler M, Paulsen P (2017) Combinatorial first-price auctions:
  Theory and experiments. Technical report, Working Paper. Technical University
  of Munich.

\bibitem[{Li(2017)}]{li2017obviously}
Li S (2017) Obviously strategy-proof mechanisms. \emph{American Economic
  Review} 107(11):3257--87.

\bibitem[{McAfee et~al.(1989)McAfee, McMillan, \protect\BIBand{}
  Whinston}]{mcafee1989multiproduct}
McAfee RP, McMillan J, Whinston MD (1989) Multiproduct monopoly, commodity
  bundling, and correlation of values. \emph{The Quarterly Journal of
  Economics} 104(2):371--383.

\bibitem[{Milgrom(2004)}]{milgrom2004putting}
Milgrom P (2004) \emph{Putting auction theory to work} (Cambridge University
  Press).

\bibitem[{Milgrom(2007)}]{Milgrom2007PackageAuctionsAndExchanges}
Milgrom P (2007) Package auctions and exchanges. \emph{Econometrica}
  75(4):935--965.

\bibitem[{Nisan(2006)}]{Nisan2006BiddingLanguagesForCombinatorialAuctions}
Nisan N (2006) Bidding languages for combinatorial auctions. Cramton P, Shoham
  Y, Steinberg R, eds., \emph{Combinatorial Auctions} (MIT Press).

\bibitem[{Nisan \protect\BIBand{} Segal(2006)}]{nisan2006communication}
Nisan N, Segal I (2006) The communication requirements of efficient allocations
  and supporting prices. \emph{Journal of Economic Theory} 129(1):192--224.

\bibitem[{Pycia \protect\BIBand{} Troyan(2019)}]{pycia2019theory}
Pycia M, Troyan P (2019) A theory of simplicity in games and mechanism design .

\bibitem[{Rabinovich et~al.(2013)Rabinovich, Naroditskiy, Gerding,
  \protect\BIBand{} Jennings}]{Rabinovich2013ComputingBNEs}
Rabinovich Z, Naroditskiy V, Gerding EH, Jennings NR (2013) Computing pure
  {B}ayesian-{N}ash equilibria in games with finite actions and continuous
  types. \emph{Artificial Intelligence} 195:106--139.

\bibitem[{Rothkopf et~al.(1998)Rothkopf, Peke\v{c}, \protect\BIBand{}
  Harstad}]{Rothkopf1998Nphard}
Rothkopf MH, Peke\v{c} A, Harstad RM (1998) Computationally manageable
  combinational auctions. \emph{Management Science} 44:1131–--1147.

\bibitem[{Sandholm(2013)}]{Sandholm2013largescale}
Sandholm T (2013) Very-large-scale generalized combinatorial multi-attribute
  auctions: Lessons from conducting \$60 billion of sourcing. Vulkan N, Roth
  AE, Neeman Z, eds., \emph{The Handbook of Market Design}, chapter~16 (Oxford
  University Press).

\bibitem[{Scheffel et~al.(2011)Scheffel, Pikovsky, Bichler, \protect\BIBand{}
  Guler}]{scheffel2011experimental}
Scheffel T, Pikovsky A, Bichler M, Guler K (2011) An experimental comparison of
  linear and nonlinear price combinatorial auctions. \emph{Information Systems
  Research} 22(2):346--368.

\bibitem[{Vickrey(1961)}]{vickrey1961counterspeculation}
Vickrey W (1961) Counterspeculation, auctions, and competitive sealed tenders.
  \emph{The Journal of Finance} 16(1):8--37.

\end{thebibliography}



%
%
%

\newpage

\ECSwitch


\ECHead{Electronic Companion}



\section{Example Auction with Natural Overbidding}
\label{app:overbiddingexample}

In this section, we provide an example of a bid that at first glance appears it should be a complex bid: it contains bundle bids on two bundles such that the marginal bid on the larger bundle exceeds the marginal valuation of that bundle.
However, upon closer inspection, it is actually a very reasonable way for a bidder to behave.
As the present paper focuses on describing complex bidding manipulations in the most crisp and clear way possible, we consider such edge cases to be simple bids, and we leave a detailed exploration of the boundary between complex bidding and other forms of overbidding to future work.

\begin{example}
Consider an auction with two bidders and two goods, and the first price payment rule.
Bidder 1 has the following valuation:
\begin{align*}
    v_1(\{1\}) = 6,\\
    v_1(\{1,2\}) = 9.
\end{align*}
Bidder 2 makes a bid that has equal probability of taking on each of two possible realizations:
\begin{equation*}
    B_2 = \left\{ \begin{array}{ll} (\{2\}, 4) & 50\% \\ (\{1,2\}, 6) & 50\% \\\end{array} \right.
\end{equation*}
\end{example}
The bid $B_2$ creates two possible worlds, and depending which of those worlds is realized, bidder 1 needs to win a different bundle to maximize her profit. The example is set up in such a way that the larger bundle is won with a narrower profit margin than the smaller one.
Thus, bidder 1's best response to $B_2$ requires a marginal overbid on good 2 given good 1, i.e., the increase in the bid when going from bundle $\{1\}$ to bundle $\{1,2\}$ is larger than the corresponding increase in value.
We formalize this reasoning below.

\begin{proposition}
    Bidder 1's best response to $b_2$ is a bid $b_1$ for which it holds that
    \begin{equation}
        b_1(\{1,2\}) - b_1(\{1\}) >  v_1(\{1,2\}) - v_1(\{1\}) = 3.
        \label{eq:overbid}
    \end{equation}

\end{proposition}

\proof{Proof.}
Consider the bid
\begin{equation*}
    b_1^* = (\{1\}, 2 + 2 \varepsilon) \oplus (\{1,2\}, 6 + \varepsilon).
\end{equation*}
It is easy to see that this bid fulfills \eqref{eq:overbid}, and it achieves an expected utility of 
\begin{equation*}
    \frac{4 + 3 - 3 \varepsilon}{2} \to 3.5.
\end{equation*}

Now, let $b_1$ be any other bid for which  \eqref{eq:overbid} does not hold. 
We show by case split that $b_1$ achieves strictly less expected utility than $b_1^*$.

Case 1: $b_1(\{1,2\}) < 6$. The expected utility is at most 3: under the first realization of $b_2$, the maximum possible utility (VCG payoff) is 6 for bundle $\{1\}$ and 5 for bundle $\{1,2\}$. Under the second realization of $b_2$, $b_1$ does not win and the utility achieved is 0.

Case 2: $b_1(\{1,2\}) \geq 6$. This implies that $b_1(\{1\}) \geq 3$ due to \eqref{eq:overbid} not holding.
Thus, the profit target (value - bid) of $b_1$ is at most 3, and the expected utility is also at most 3.
\Halmos
\endproof


\section{BNE proofs for Example~\ref{ex:1}}
\label{app:BNEproof}

\begin{lemma}
The strategy profiles given in Table~\ref{tab:example} are a simple and complex BNE, respectively.
\end{lemma}

\proof{Proof.}
Bidder 3 has a dominant strategy to bid truthfully \citep{beck2013incentives}.
In both BNEs, bidder 2's expected utility is 
\begin{equation}
\frac{1}{2} \cdot \left(\frac{4}{5} - \frac{1}{2} \right) = 0.15
\end{equation}
Any bid lower than $0.5$ would reduce the probability of winning (and thus the utility) to 0.
Bids strictly between $0.5$ and $1$ would increase the payment per Lemma~\ref{lem:vcgn_simple}.
A bid of $1.0$ or more would make bidder 2 win even when bidder 1 has low valuation, but in this case bidder 2's payment would be above her value, further decreasing total utility.
In the case of a complex bid, a bundle bid on $\{1,2\}$ below $1$ never wins and does not affect the payments, and at $1$ or above it again leads to negative utility.

The argument for bidder 1 is identical for the simple BNE, except when the bid is at least $1$, where we explicitly compute the expected utility to be 
\begin{equation}
    \frac{1}{2} \cdot \left(\frac{6}{5} - \frac{3}{4} \right) + \frac{1}{2} \cdot \left(\frac{6}{5} - 1 \right) = 0.325.
\end{equation}
which is smaller than the utility of 
\begin{equation}
    \frac{1}{2} \cdot \left(\frac{6}{5} - \frac{1}{2} \right) = 0.35
\end{equation}
obtained under the equilibrium bid.
For the complex BNE, it can easily be checked that bidder 1's payment is identical to her VCG payment, so she captures the highest utility possible under any core-selecting payment rule.
\Halmos

\endproof


\section{Characterization of VCG-nearest payments in LLG}
\label{app:familyVCGN}



\begin{lemma}
Assume that the global bidder makes a simple bid with $b_3(\{1,2\}) = \bar{\beta}$.
Also assume that each local bidder $i$ makes a simple bid.
Then, if the local bidders win, the VCG-nearest payment of local bidder $i$ is
\begin{equation}
\vcgn_i = \frac{1}{2} \cdot \left(\vcg_i + \min(b_i(\{i\}), \bar{\beta}))   \right).
\end{equation}
\label{lem:vcgn_simple}
\end{lemma}

\proof{Proof.}
It is easy to check that
\begin{equation}
    \vcg_j := \max(0, \bar{\beta} - b_i(\{i\})) = \bar{\beta} - \min(b_i(\{i\}), \bar{\beta}).
\end{equation}

The main core constraint is the one imposed by the global bidder, requiring the joint payment of the locals to be at least $\bar{\beta}$:
\begin{equation}
    p_1 + p_2 \geq \bar{\beta}.
    \label{eq:global_core}
\end{equation}
The extra amount that the locals need to jointly pay on top of VCG is 
$$\bar{\beta} - \vcg_i - \vcg_j = \min(b_i(\{i\}), \bar{\beta}) - \vcg_i.$$
Even if the whole missing amount was paid by bidder $i$, the payment would be $\min(b_i(\{i\}), \bar{\beta})$ which still falls inside $i$'s IR constraint.
The only other core constraints are the VCG constraints and a pair of constraints weaker than VCG.
Thus, the point on the MRC closest to VCG is the one that increases payments from VCG evenly for both local bidders, and the VCG-nearest payment is:
\begin{align}
    \label{eq:VCGN_appA}
    \vcgn_i &= \vcg_i 
    + \frac{1}{2} \cdot \left(\min(b_i(\{i\}), \bar{\beta}) - \vcg_i   \right) \\
           &= \frac{1}{2} \cdot \left(\vcg_i + \min(b_i(\{i\}), \bar{\beta}))   \right).
\end{align}
\Halmos

\endproof

\begin{lemma}
Assume that the global bidder makes a simple bid with $b_3(\{1,2\}) = \bar{\beta}$, and local bidder $j$ makes a simple bid.
%
If bidder $i$ makes a bid of the form
$$ b_i = (\{i\}, \beta + \epsilon) \oplus (\{1,2\}, \bar{\beta} + \epsilon),$$
for $0 \leq \beta < \bar{\beta}$ and $ \epsilon \geq 0$, then if both local bidders win, the VCG-nearest payment of local bidder $i$ is
\begin{equation}
\vcgn_i = \frac{1}{2} \cdot \left(\vcg_i + \min(b_i(\{i\}), \bar{\beta}))   \right).
\end{equation}
\label{lem:vcgn_complex}
\end{lemma}

%

\proof{Proof.}
The core constraints involving bidder $i$ on the blocking side are:
\begin{align}
p_j       &\geq  (\bar{\beta} + \epsilon) - (\beta + \epsilon) \geq \bar{\beta} - \beta, \\
p_j + p_3 &\geq  (\bar{\beta} + \epsilon) - (\beta + \epsilon) \geq \bar{\beta} - \beta, \\
p_3       &\geq  0,
\end{align}
where $j$ is the other local bidder.
We note that the first of these constraints (which corresponds to bidder $j$'s VCG payment) is the only relevant one, since $p_3 \leq 0$ by IR.
If we set $\epsilon = 0$, bidder 1's second atomic bid could be removed without changing that first constraint, and thus we can apply Lemma~\ref{lem:vcgn_simple} to establish the $VCGN$ payments as claimed.
To extend the result to $\epsilon > 0$, note that both the shape of the core and the VCG payments are independent of $\epsilon$, and thus the $VCGN$ payments must also be independent of $\epsilon$, since they are calculated only based on those two things.
\Halmos
\endproof
\section{Proof of Theorem~\ref{thm:VCG-nearest}}
\label{app:VCG-nearest-proof}

\todo{go over this proof, harmonize structure and notation with the FP version}

\para{Intro + setup}
The main difference between the proof of Theorems~\ref{thm:FP} and~\ref{thm:VCG-nearest} is that we construct a more complicated auction family with more bidders, to make sure that the VCG-nearest payment behaves identically to first-price for bidder 1.
This is achieved by moving the core far away from the VCG payments of other bidders, causing the projection onto the minimum revenue core to prioritize reducing the payments of those other bidders over that of bidder 1.
Lemma~\ref{lem:vcg-nearest-payments} (the analogue of Lemma~\ref{lem:fp-payments}) must take into account many core constraints to characterize bidder 1's payments.


\para{Auction Family}
We adapt the family of auction instances $F(m+1)$ used in the previous section, as follows:

\begin{definition}
Let $m$, $\mathcal{L}$, $\sigma$, $f(i)$, $C$ and $j$ be given as in Definition~\ref{def:fp-family}.
Then, the auction $F'(m+3)$ is given as follows:
\begin{enumerate}
\item The auction has $5$ bidders and $m + 3$ goods.
\item Bidder 1 has value $1$ for bundle $\{m+1\}$.
\item Bidders 2 and 3 always bid $C$ on bundles $\{m+2\}$ and $\{m+3\}$, respectively.
\item Bidder 4 always bids $C$ on bundle $\{m+1, m+2, m+3\}$.
\item Bidder 5 bids $C$ on bundle $\sigma^{-1}(j)$, and $C + f(j)$ on bundle $\sigma^{-1}(j) \cup \{m+1\}$.
\item The allocation rule is efficient. For ease of presentation, ties are broken in favor of bidder 1 when relevant.
\item The payment rule is VCG-nearest.
\end{enumerate}
\end{definition}

The bidders of this auction family and the relation of their bundles to each other are shown in Table~\ref{tab:vcg-nearest}.
The first three bidders are in direct competition with bidder 4, and outbid her easily.
However, bidder 1 must now also outbid bidder 5, for which she needs an atomic bid on an active bundle of at least $f(j)$.

\begin{figure}
\FIGURE
{
\includegraphics[width=0.85\textwidth]{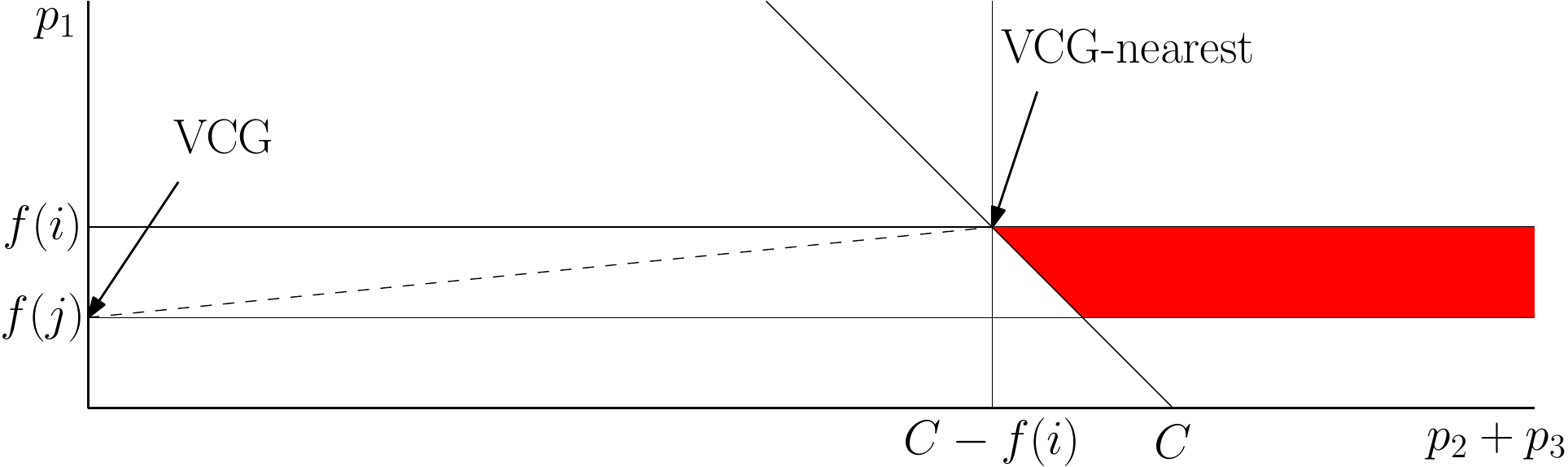}
}
{
The core in auction family $F'(m+3)$.
Bidders 2 and 3 are symmetrical, so we show their joint payment on one axis. Since VCG-nearest minimizes $L_2$ distance to VCG, it is easy to see that the optimal solution is to charge bidder 1 as much as the individual rationality constraint will allow.
\label{fig:core}
}
{}
\end{figure}

The shape of the core is depicted in Figure~\ref{fig:core}, showing that bidder 1's payments are identical to first-price.
The proof proceeds similarly to the first-price case.
    

\begin{table}
\TABLE
{
Bundle bids made by different bidder types in auction family $F'(m+3)$.
Bidder 1 wins her desired good $m+1$ whenever $i \geq j$, i.e., she outbids the competitor. Bidder 1 can make additional bundle bids that include more goods, and those bids are active whenever they do not overlap with $\sigma^{-1}(j)$.
\label{tab:vcg-nearest}
}
{
\begin{tabular}{lc||cccc}

\multicolumn{2}{c||}{\bf Bidders} &
\multicolumn{4}{c}{\bf Goods} \\

&&
\bf Signalling Goods & 
\bf Prize &
\multicolumn{2}{c}{\bf Helper Goods} \\

&&
$1$
\ldots
$m$ &
$m+1$ & 
$m+2$ &
$m+3$ \\
\hhline{=:=::=:=:=:=}

\bf Protagonist &1 &  & \multicolumn{1}{|c|}{$f(i)$} & &\\
\cline{1-2}
\cline{4-5}

\multirow{3}{1cm}{\bf Helper Bidders} &2& \multicolumn{2}{c|}{}& \multicolumn{1}{c|}{$C$} & \\
\cline{5-6}
&3& \multicolumn{3}{c|}{}& \multicolumn{1}{c|}{$C$} \\
\cline{4-6}

&4 & & \multicolumn{3}{|c|}{$C$} \\
\cline{1-2}
\cline{3-6}

\cline{6-6}
\multirow{2}{1cm}{\bf Competitor} & \multirow{2}{0.1cm}{5} & \multicolumn{1}{c|}{$C$ on $\sigma^{-1}(j)$} &  &\\
\cline{3-4}
& & \multicolumn{2}{c|}{$C + f(j)$ on $\sigma^{-1}(j) \cup \{m+1\}$ }& &\\
\cline{3-4}

\end{tabular}
}
{}
\end{table}

\begin{lemma}
    In the auction family $F'(m+3)$, let $b_1$ be a bid for bidder 1, and let $K^*$ be the bundle with the highest active bid in $b_1$.
    Furthermore, assume that bidder 1's bids on other bundles are not too high, i.e., $\forall K \ne K^*: b_1(K) \leq 1$.
    
    Then, bidder 1 wins bundle $K^*$ if $b_1(K^*) \geq f(j)$, and her payment is
    \begin{equation}
        \min\left(b_1, \frac{C + 2f(j)}{3}\right).
    \end{equation}
    Otherwise, bidder 1's allocated bundle does not include good $m+1$.
    \label{lem:vcg-nearest-payments}
\end{lemma}

\proof{Proof.}
    Because bidder 1's bids are capped at $1$, she can never win with a bid that is blocked by bidders 2,3 or 5.
    It is clear that bidders 2 and 3 always win their respective bundles, and bidder 4 never wins.
    It follows that good $m+1$ is won by one of the remaining two bidders.
    Bidder 1 outbids bidder 5 exactly when she has an active bundle on which she bids above $f(j)$.

    Next, we derive the core constraints (and the VCG payments).
    There are 5 bidders and thus $2^5 - 1 = 31$ core constraints.
    Since bidder 4 always loses, any constraint corresponding to blocking coalition $L$ with $4 \notin L$ is dominated by the constraint of coalition $L \cup \{4\}$.
    This accounts for 16 core constraints, leaving 15.
    
    We next deal with the remaining constraints that involve bidder 5 on the left hand side.
    Let $K'$ be the bundle with the highest bid of bidder 1 (not necessarily active).
    Then, the 8 constraints are:
    \begin{align}
        p_5 &\geq 2C + b_1(K') - 2C - b_1(K^*) = b_1(K') - b_1(K^*),\\
        p_1 + p_5 &\geq 2C - 2C = 0,\\
        p_2 + p_5 &\geq C + b_1(K') - C - b_1(K^*) = b_1(K') - b_1(K^*),\\
        p_3 + p_5 &\geq C + b_1(K') - C - b_1(K^*) = b_1(K') - b_1(K^*),\\
        p_1 + p_2 + p_5 &\geq C - C = 0,\\
        p_1 + p_3 + p_5 &\geq C - C = 0,\\
        p_2 + p_3 + p_5 &\geq C - b_1(K^*),\\
        p_1 + p_2 +p_3 + p_5 &\geq C - 0 = C.
    \end{align}
    Of these constraints, the first one is the VCG constraint, which dominates the 3rd and 4th constraints. The 7th and 8th constraints are dominated by constraints introduced below, and the rest are 0.
    
    The remaining 7 core constraints are:
    \begin{align}
        p_1 &\geq 3C + f(j) - 3C = f(j),\\
        p_2 &\geq 2C + b_1(K^*) - 2C - b_1(K^*) = 0,\\
        p_3 &\geq 2C + b_1(K^*) - 2C - b_1(K^*) = 0,\\
        p_1 + p_2 &\geq 2C + f(j) - 2C = f(j),\\
        p_1 + p_3 &\geq 2C + f(j) - 2C = f(j),\\
        p_2 + p_3 &\geq 2C - C - b_1(K^*) = C - b_1(K^*),\\
        p_1 + p_2 + p_3 &\geq 2C - C = C.
    \end{align}
    The first of these is bidder 1's VCG constraint, which dominates the 4th and 5th.
    
    In summary, the VCG payment vector is as follows:
    \begin{align}
        \vcg_1 &= f(j),\\
        \vcg_2 &= 0,\\
        \vcg_3 &= 0,\\
        \vcg_4 &= 0,\\
        \vcg_5 &= b_1(K') - b_1(K^*),
    \end{align}
    and the relevant core constraints (other than VCG and IR constraints) are:
    \begin{align}
        p_2 + p_3 &\geq C - b_1(K^*),\\
        p_1 + p_2 + p_3 &\geq C.
    \end{align}
    
    We now derive the VCG-nearest payments.
    It is clear that bidder 5's VCG-nearest payment is her VCG payment, since any increase in payment does nothing to help fulfill any of the other constraints.
    Furthermore, the VCG-nearest payment must be symmetric for bidders 2 and 3, since they only occur together in any constraint that is not dominated by another constraint.
    Any asymmetric payment vector in the core could be brought closer to VCG by balancing $p_2$ and $p_3$, without violating any constraints.
    Therefore, the VCG-nearest payment can be characterized by a single parameter $\Delta$ as follows:
    \begin{align}
        p_1 &= \Delta,\\
        p_2 &= \frac{C - \Delta}{2},\\
        p_3 &= \frac{C - \Delta}{2},\\
        p_4 &= 0,\\
        p_5 &= b_1(K') - b_1(K^*).
    \end{align}

    The distance from $\vcg$ to $p$ in terms of $\Delta$ is
    \begin{equation}
        \left\| p - \vcg \right\|^2 = (\Delta - f(j))^2 + 2\cdot\left(\frac{C - \Delta}{2}\right)^2.
    \end{equation}
    Taking the derivative w.r.t. $\Delta$, we get
    \begin{equation}
        \frac{d}{d \Delta} \left\| p - \vcg \right\|^2 = 2(\Delta - f(j)) - (C - \Delta) = 3\Delta -2f(j) - C,
    \end{equation}
    so the distance is minimized at
    \begin{equation}
         \Delta = \frac{2f(j) + C}{3}.
    \end{equation}
    When $b_1(K^*)$ is below this threshold, this payment is outside the core (due to bidder 1's IR constraint).
    In that case, the derivative is negative over the whole domain $[0, b_1(K^*)]$, so the optimal solution is $\Delta = b_1(K^*)$.
    \Halmos

\endproof

For this particular auction, bidder 1's payment is a function of only $j$, independent of the identity of $K^*$, just like in the first-price case.
Computing bidder 1's simple and complex best response utilities is now completely analogous to what we did in the previous section.

\begin{lemma}
    In auction family $F'(m+3)$, bidder 1's simple best response utility is $\displaystyle \Theta\left(\frac{1}{2^{2^m}}\right)$.
    \vspace{2mm}
    \label{lem:singleminded_brutil}
\end{lemma}

\proof{Proof.}
    Analogous to Lemma~\ref{lem:fp_singleminded_brutil}, substituting
    Lemma~\ref{lem:fp-payments} with Lemma~\ref{lem:vcg-nearest-payments}.
    \Halmos
\endproof

\begin{lemma}
    In auction family $F'(m+3)$, bidder 1's best response utility is $\displaystyle \Theta\left(\frac{2^m}{2^{2^m}}\right)$.
    \vspace{2mm}
    \label{lem:full_brutil}
\end{lemma}

\proof{Proof.}
    Analogous to Lemma~\ref{lem:fp_full_brutil}, substituting
    Lemma~\ref{lem:fp-payments} with Lemma~\ref{lem:vcg-nearest-payments}.
    \Halmos
\endproof

\proof{Proof of Theorem~\ref{thm:VCG-nearest}.}
    Consider the family of auctions $F'(m+3)$.
    The total number of goods is $m + 3 = \theta(m)$.
    Using Lemmas~\ref{lem:singleminded_brutil} and~\ref{lem:full_brutil}, the ratio of the simple and complex best response utilities of bidder 1 is
    \begin{equation}
           \Theta\left( \frac{1}{2^{2^m}} \middle/ \frac{2^m}{2^{2^m}}  \right)
        = \Theta\left( \frac{1}{2^m} \right).
    \end{equation}
    \Halmos
\endproof
\vspace{-3mm}


\end{document}